\documentclass[11pt,letterpaper]{article}

\usepackage{amsmath}
\usepackage{amsfonts}
\usepackage{amssymb}
\usepackage[pdftex]{graphicx}
\usepackage{verbatim}
\usepackage{authblk}
\usepackage{url}
\usepackage{wrapfig}

\newtheorem{con}{Conjecture}

\newtheorem{defin}{Definition}

\newtheorem{lem}{Lemma}

\newtheorem{prop}{Proposition}
\newtheorem{rem}{Remark}
\newtheorem{theo}{Theorem}

\newtheorem{nullhypothesis}{Null Hypothesis}

\makeatletter
\newcommand\ackname{Acknowledgements}
\if@titlepage
  \newenvironment{acknowledgements}{%
      \titlepage
      \null\vfil
      \@beginparpenalty\@lowpenalty
      \begin{center}%
        \bfseries \ackname
        \@endparpenalty\@M
      \end{center}}%
     {\par\vfil\null\endtitlepage}
\else
  \newenvironment{acknowledgements}{%
      \if@twocolumn
        \section*{\abstractname}%
      \else
        \small
        \begin{center}%
          {\bfseries \ackname\vspace{-.5em}\vspace{\z@}}%
        \end{center}%
        \quotation
      \fi}
      {\if@twocolumn\else\endquotation\fi}
\fi
\makeatother

\begin{document}

\title{\textsc{Sandpile cascades on interacting tree-like networks}}

\author[1,2,*]{Charles D. Brummitt}
\author[2,3,4]{Raissa M. D'Souza}
\author[5,6]{E. A. Leicht}

\affil[1]{\small Department of Mathematics, University of California, Davis, CA, 95616}
\affil[2]{Complexity Sciences Center, University of California, Davis, CA, 95616}
\affil[3]{Dept. of Mechanical and Aeronautical Engineering and Dept. of Computer Science, University of California, Davis, CA, 95616}
\affil[4]{The Santa Fe Institute, Santa Fe, NM 87501}
\affil[5]{CABDyN Complexity Centre, University of Oxford, Oxford, United Kingdom}
\affil[6]{Sa\"{i}d Business School, University of Oxford, Oxford, United Kingdom}

\date{\today}

\maketitle
\thispagestyle{empty}

\let\oldthefootnote\thefootnote
\renewcommand{\thefootnote}{\fnsymbol{footnote}}
\footnotetext[1]{To whom correspondence should be addressed. Email: cbrummitt@math.ucdavis.edu}
\let\thefootnote\oldthefootnote

\vspace{-1cm}
\begin{abstract} 
The vulnerability of an isolated network to cascades
 is fundamentally affected by its interactions with other networks. Motivated by failures cascading among electrical grids, we study the Bak-Tang-Wiesenfeld sandpile model on two sparsely-coupled random regular graphs.
  By approximating avalanches (cascades) 
   as a multi-type branching process and using a generalization of Lagrange's expansion to multiple variables, we calculate the distribution of avalanche sizes within each network. 
  Due to coupling, large avalanches in the individual networks are mitigated---in contrast to the conclusion for a simpler model \cite{NewmanHICSS04}. Yet when compared to uncoupled networks, interdependent networks more frequently suffer avalanches that are large in both networks. Thus sparse connections between networks stabilize them individually but destabilize them jointly, as coupling introduces reservoirs for extra load yet also inflicts new stresses. 
  These results suggest that in practice, to greedily mitigate large avalanches in one network, add connections between networks; conversely, to mitigate avalanches that are large in both networks, remove connections between networks.  We also show that when only one network receives load, the largest avalanches in the second network increase in size and in frequency, an effect that is amplified with increased coupling between networks and with increased disparity in total capacity. Our framework is applicable to modular networks as well as to interacting networks and provides building blocks for better prediction of cascading processes on networks in general.
\end{abstract}


\section{Introduction}
The interdependence among systems is paramount.
A system that is stable in isolation, for instance, may lose stability when coupled to another system. By contrast, some systems are largely useless unless coupled to others; important examples of this occur in biology, in which modular components---organs, tissues, cells and organelles---function properly only in churning, interdependent synchrony. Such interdependence among systems can be detrimental, however, as small failures can escalate to catastrophe. Just one failed element within one module of a system---a malignant lymph node in a human body, an eradicated autotroph in an ecosystem, a downed power line in an electrical grid---can spread to other connected subsystems and, via feedback loops and percolation on a complex web of links, cascade to system-wide failure. The connections among systems can enhance or inhibit, 
hasten or delay such cascades.  

One of the best examples of coupled systems prone to cascading failures is infrastructure~\cite{Little2002,Rinaldi2004,INLsurvey,Panzieri2008}. There is no denying the increasing interdependence of modern infrastructure: water, gas, Internet packets, financial transactions, phone calls and electrons move on regional, national and global networks, responding to demand and depending on one another for proper function. When an element of one of these networks is overwhelmed, its neighboring elements may pick up the slack---they are often engineered to do so---but when they cannot, the failure can spread, both within that network and to others.

Striking examples of failures cascading among interdependent infrastructure abound. In 1998, for instance, a telecommunications satellite over the western United States malfunctioned, crippling the communication network, which in turn disrupted the transportation network because gas stations could not process credit card transactions and airports lacked precise weather information \cite{Panzieri2008}. In Italy in 2004, the failure of a telecom node paralyzed nearly all national telecommunication services, in turn halting most Italian financial transactions, postal deliveries and flights \cite{Panzieri2008}. More recently, the ash-covered skies over Europe in the aftermath of the eruption of the Eyjafjallaj\"okull volcano in Iceland on April 14, 2010, halted nearly all flights in western Europe, which not only disrupted cultural, sporting, military and diplomatic events but also affected economies across the globe due to their unprecedented interdependence. It is remarkable the extent to which failures and other information ripple among today's increasingly intertwined economies, penetrating the far corners of the globe: thousands of flower farmers in Kenya, for instance, were laid off because their harvests could not be shipped by air to the UK in the aftermath of Eyjafjallaj\"okull, layoffs that in turn affected the Kenyan economy \cite{volcano_dailymail}.

Such interdependence motivates analytic models of coupled networks. In these models, each infrastructure system is a network, yet nodes in one network may depend on nodes in other networks. For example, the World Wide Web is a virtual network that runs on the physical Internet of routers, which in turn depends on the electrical grid for power, all three of which are (increasingly) crucial to the finance, transportation, health and other systems. The failure of nodes in one network may cause nodes in other networks to fail, which in turn impairs nodes in other networks, and so on. This recursive structure of the failures hints at the possibility of rigorous, mathematical analysis (e.g., \cite{buldyrev}). 
A better understanding of the characteristics of coupled networks that facilitate or inhibit spreading processes on them may help to answer the difficult question of design: How do we fortify interdependent infrastructure against catastrophe? Furthermore, how can we effectively spread desirable ``pathogens'', such as information and ideas, among interdependent networks?

The idea of interacting networks is closely related to the concept of modules or communities within networks. In the classic definition, modules are subgraphs with significantly more edges than expected within the subgraph and few edges to the rest of the network~\cite{GirvanNewmanPNAS02}. Alternate definitions of modules have been studied that capture, for example, hierarchical structure \cite{clauset_hierarchical} or similarity in patterns of connectivity~\cite{NewmanLeichtPNAS2007}. For a recent comprehensive review of community structure, see Ref.~\cite{csCompareFortunato}. Despite the extent of research on finding modules in networks, their use in applications so far has been limited. Our aim is to use knowledge of the modular structure of isolated networks (or of the structure of interactions among distinct networks) to better predict the behavior of the whole system.

The ``electrical grid'' of the United States, for example, is not a grid but instead consists of a collection of over 3,200 distinct local and regional electric utilities~\cite{eia_overview2007}.  A small number of these are federal utilities, while the vast majority are either investor-owned or publicly-owned. Each individual utility is an independent entity, with myriad economic and physical considerations dictating its structure and operating policies, yet the entities connect together to form the ``grid". Historically the grid has comprised a collection of loosely interconnected, local systems, but the level of connectedness, length scales of interactions, and the number of small, distributed power sources are increasing~\cite{aminMRS2008}. Given this increasing level of diversity and interconnectivity, it is important to understand at a fundamental level the impact of interactions on cascading failures. 

With coupled electrical grids in mind---though not limited to this application---we study the Bak-Per-Wiesenfeld (BTW) sandpile model \cite{BTW_SOC} on interacting networks. In the sandpile model, grains of sand are randomly dropped on nodes, each of which possess an innate capacity to hold sand. Whenever a node's load exceeds its capacity, it ``sheds'' its load to its neighbors, and any nodes now exceeding their capacity shed their load synchronously at the next time step. This process of unstable nodes toppling continues until equilibrium is restored. In this way, dropping a single grain of sand can cause an avalanche, often small but sometimes spanning the entire network. Sandpile models on networks are toy models that do not perfectly capture the structure and dynamics of electrical grids, but they share enough properties to be relevant by elucidating the space of possible behaviors in simple models. Moreover, sandpile models are simple enough to find applications in other systems that bear and shed load and that yield cascades with the long correlations and power-laws that characterize self-organized criticality (e.g.,  neuronal networks \cite{BeggsPlenz2003, JuanicoJPhysA2007}; forest fires, earthquakes, landslides, financial markets~\cite{bakBook}). 

There is a large body of research on cascading failures in infrastructure~(e.g., \cite{Little2002,Rinaldi2004, NewmanHICSS04,INLsurvey,dobson_chaos2007,Panzieri2008,Rosato2008}), but only recently has the effect of the interdependence among systems garnered considerable attention. Newman et al.\ have studied two such models: CASCADE, a sandpile model in which load (sand) is shed uniformly to nodes in two networks in a non-local way \cite{NewmanHICSS04}; and DSCM, a probabilistic model of random failures and repairs in which neighbors of failed nodes are more likely to fail \cite{NewmanHICSS04}. The CASCADE model, approximated by a branching process, demonstrates how the coupling of networks shifts the critical points of global cascades. However, to reach this conclusion analytically they make the unrealistic (but mathematically convenient) assumption that load is shed globally to every node rather than locally, so that critical points are simply maximal eigenvalues of the network coupling matrix. Meanwhile, DSCM (short for ``coupled dynamic complex system model'') is a simple model that exhibits self-organized criticality---i.e., the system dynamically arranges itself at the critical point, where it exhibits long time correlations and power-law distributions of failure size. Newman et al.\ conclude that introducing coupling between systems makes them more vulnerable, though the models are simplified and deserve more sophisticated, realistic studies.

A model due to Bulydrev et al.\ \cite{buldyrev} is also simplified for the sake of analytical solutions. They consider two networks, each with the same number of nodes, and these networks interact in that each node in one network is connected uniformly at random to one node in the other network. Initially, a random subset of nodes in one network fail, and then a cascade of failures ensues, in which internal edges that connect disconnected components (i.e., maximally connected subgraphs) in the other network are deleted. Although the assumptions regarding the network coupling are somewhat unrealistic (every node has one neighbor in the other network, chosen uniformly at random) and although the failure mechanism is very specific (delete internal edges that connect disconnected components), the model nevertheless demonstrates the surprising catastrophes possible in interdependent networks. Moreover, considering the interdependence between networks can upend longstanding notions about isolated networks. For example, Buldyrev et al.\ found that for coupled networks with their failure mechanism, broader degree distributions are more vulnerable to random failure \cite{buldyrev}, in contrast to the conclusion for isolated networks that broader degree distributions are more robust to random failure \cite{error_attack_albert_barabasi}.

In this paper we analyze cascades in sandpiles on interacting networks by approximating them by multi-type branching processes. Here we consider locally tree-like networks, as are the case for random graphs generated using the configuration model as in Ref.~\cite{watts_newman_strogatz}. Studies of the electrical grid suggest that they are tree-like (low clustering coefficient) and have narrow degree distribution: two small-scale studies of the topology of nodes constituting a small regional electric grid suggest that the connectivity of nodes obeys an exponential distribution~\cite{amaralPNAS2000,albertPRE2004}, and a recent extensive study corroborates the exponential nature of the connectivity as well as a small non-zero clustering coefficient ($C=0.071$)~\cite{Hines2010}.  The framework we develop is for arbitrary degree distribution but requires tree-like graphs (with $C\approx 0$). Extensions to social and other networks that are \emph{not} tree-like would build off recent work on random graphs with arbitrary subgraphs \cite{newman_subgraphs, miller_percolation, newman_clustering}. We solve the multi-type branching processes for the avalanche size distributions using an old result in the mathematics literature, a generalization of Lagrange's expansion due to I. J. Good \cite{good_lagrange}, as well as numerically simulate a situation not yet described by our analytic framework: asymmetric load applied to one network rather than to both. By studying processes on modular or interacting networks using multi-type branching processes and by using knowledge about the connection structure within and between modules (as in \cite{gleeson_cascademodular, vazquez_heterogeneouspopulations, vazquez_structuredpopulations,DoroMenPRE08,OstilliMendesPRE09,PourboPRE09}), we begin to develop a more complete picture of the behaviors of complex, heterogeneous networks.

We show that introducing coupling between networks stabilizes them individually, in the sense that large avalanches are mitigated and small avalanches are amplified, yet destabilizes them jointly, in that avalanches that are large in both networks become more likely---compared to the null hypothesis of two uncoupled graphs. The finding that connecting networks stabilizes them individually contrasts to the result in \cite{NewmanHICSS04} for a simpler sandpile model, and it assuages the warnings in \cite{buldyrev} about the catastrophic cascades of failed connectivity in a model of coupled networks. On the other hand, we also find that introducing connections between networks can destabilize them individually in some circumstances: if only one network receives external load, the other network suffers large avalanches that increase in severity and frequency with increased coupling between networks and with increased disparity in relative capacity. This suggests an arms race to increase capacity of a network to fortify against cascades inflicted by neighboring networks.

\subsection{Classic BTW sandpile models on lattices}
Introduced in the late 1980s \cite{BTW_SOC}, the Bak-Tang-Wiesenfeld sandpile model is a well-studied toy model of cascades that exhibits self-organized criticality, power laws and universality classes. In a classic version of the model on a finite, two-dimensional lattice, grains of sand are dropped uniformly at random on nodes in the lattice, and whenever a node contains four or more grains of sand, it sheds one grain to each of its four neighbors at the next time step. The lattice has open boundaries, so that sand shed off the boundary of the lattice is lost, which prevents the system from becoming inundated with sand. Various measures of the size, area, and duration of avalanches follow power-laws, and these variables relate to one another via power laws \cite{benhur}.

A few variants of the sandpile model on lattices can be solved exactly if the shedding rules have abelian symmetry \cite{benhur}. More recently, sandpile models have been studied on (isolated) networks, including Erd\H{o}s-R\'{e}nyi graphs \cite{sandpile_ER, sandpile_ER_lise}, scale-free graphs \cite{goh_sandpile_powerlaw}, and graphs generated by the Watts-Strogatz model on one-dimensional \cite{sandpile_wattsstrogatz_1d} and on two-dimensional \cite{sandpile_wattsstrogatz_2d} lattices. A common question explored using asymptotic calculations and computer simulation is: Under what conditions on the network structure is the avalanche behavior ``mean-field''---i.e., approximately that of a complete graph---which corresponds to the avalanche size distribution being a power law with exponent 3/2.

\subsection{BTW sandpile models on arbitrary networks}
The most natural choice for the ``thresholds'' or ``capacities'' for each node in a network is its degree, so that nodes shed one grain to each neighbor (or, more precisely, one grain along each outgoing edge, since it may have parallel edges or self-loops) \cite{goh_sandpile_powerlaw}. Another choice for sand thresholds is \emph{uniform} \cite{goh_sandpile_powerlaw, sandpile_ER_lise}: for example, leaves (i.e., nodes with degree one) have threshold one, while all other nodes have threshold two. However, uniform threshold suffers from two drawbacks: shedding becomes ambiguous because nodes must shed to a random subset of their neighbors rather than one grain to every neighbor, and a large portion of the novelty of sandpiles on networks (rather than on lattices) is the disparity among nodes in the amount of sand shed. As a result, here we choose thresholds of nodes to be their degrees.

To make this explicit, the sandpile dynamics on networks are as follows. With the graph fixed, we add grains of sand to nodes in the network chosen uniformly at random. Whenever the sand on a node exceeds its total degree, the node ``topples'' and sheds one grain to each of its neighbors. (More precisely, it sheds one grain along each edge, so if it has multiple edges to a neighbor, then it sends as many grains to that node, but as discussed below in Section \ref{Subsection:generating_networks_from_degree_distribution}, parallel edges in the sparse random graphs considered here are rare.) If any new nodes become unstable, they all shed synchronously at the next time step. (If a node has \emph{strictly} more grains than its degree, it still sheds one grain to each neighbor, and the leftover grains remain on the node; in our studies, the chance that a node has at least twice as many grains as its degree---and hence has to shed multiple grains per neighbor---is negligibly small.) Topplings continue until no node exceeds its capacity---i.e., equilibrium is restored---whereupon the process repeats.

To ensure that the network doesn't become overloaded with sand, whenever a node sheds its sand, each grain is deleted independently with probability $f$, which we call the \emph{dissipation rate of sand}. The dissipation of sand in sandpile models on networks is the analogy of open boundary conditions in classic BTW sandpile models on finite lattices, in which grains shed off the boundary are lost. This parameter can profoundly affect the duration and size of avalanches. In \cite{goh_sandpile_powerlaw}, they used $f = 0.0001$ for a scale-free network with $10^4$ nodes, but they did not mention the sensitivity of the avalanche size on $f$: decreasing $f$ prolongs avalanches, which can add a ``hump'' to the avalanche size distributions at large avalanches. We discuss how to choose $f$ in Section \ref{Subsection:shedding_branch} below.

\section{Random graph model of two coupled networks}
Here we study the sandpile model on two interacting networks, labeled $a$ and $b$, which have their own internal (or intra-)degree distribution and which are sparsely coupled by edges, called inter-edges. (Throughout we use the prefixes \emph{intra-} and \emph{inter-} to refer to edges and degrees \emph{within} or \emph{between} the two networks, respectively.) The connectivity of interacting networks was studied in \cite{leichtdsouza}. Each node has an intra- and inter-degree, which are the numbers of neighbors within its network and in the other network; we choose the thresholds of sand to be the most natural one, the total degree.

\subsection{Inter- and intra-degree distributions}
Since degrees play a key role in avalanche dynamics, we study in detail the degree distributions of interacting networks. Each node has an integer number $k_a \geq 0$ many $a$-neighbors and $k_b \geq 0$ many $b$-neighbors. Each network, $a$ and $b$, is characterized by a degree distribution on $\mathbb{Z}_{\geq 0}^2$:
\begin{align}
p_a(k_a, k_b) &\equiv \text{fraction of $a$-nodes with $k_a$ $a$-neighbors and $k_b$ $b$-neighbors}, \\
p_b(k_a, k_b)&\equiv \text{fraction of $b$-nodes with $k_a$ $a$-neighbors and $k_b$ $b$-neighbors}.
\end{align}
The associated generating functions are
\begin{align*}
G_a(\omega_a, \omega_b) &= \sum_{k_a,k_b = 0}^\infty p_a(k_a, k_b) \omega_a^{k_a} \omega_b^{k_b},\\
G_b(\omega_a, \omega_b) &= \sum_{k_a,k_b = 0}^\infty p_b(k_a, k_b) \omega_a^{k_a} \omega_b^{k_b},
\end{align*}
for $\omega_a, \omega_b \in \mathbb{C}$ with $|\omega_a|, |\omega_b| \leq 1$ (i.e., in the bidisc).

\begin{defin}We say that the marginal $p_{aa}(\cdot) = \sum_{k_b = 0}^\infty p_a(\cdot, k_b)$ is the \emph{intra-degree distribution} of network $a$, while the marginal $p_{ab}(\cdot) = \sum_{k_a = 0}^\infty p_a(k_a, \cdot)$ is the \emph{inter-degree distribution} of network $a$. The \emph{intra-} and \emph{inter-degree distributions} of $b$ ($p_{ba}$ and $p_{bb}$) are defined analogously. \end{defin}
It is convenient to assume that the intra- and inter-degrees of nodes are independent, but this is rarely the case in real-world networks; the connections between different infrastructures may occur most frequently between nodes of low degree, for example.
\begin{defin}The intra- and inter-degree distributions are \emph{independent} if the degree distributions can be written as a product of two probability distributions on $\mathbb{Z}_{\geq 0}$: e.g.,
\begin{align*}
p_a(k_a, k_b) &= p_{aa}(k_a) p_{ab}(k_b),\\
p_b(k_a, k_b) &= p_{ba}(k_a) p_{bb}(k_b).
\end{align*}
\end{defin}

\begin{rem}It may be that the degree distribution of one of the networks is independent while that of the other is not.\end{rem}

\subsection{Generating interacting networks from their degree distributions}\label{Subsection:generating_networks_from_degree_distribution}

We generate coupled random graphs from their degree sequences using a simple extension of the configuration model. However, this necessarily changes the inter-degree distributions in a subtle way, due to the fact that the edges between networks are undirected. Below we make precise this effect of conditioning on the event that there must be equally many edges from $a$ to $b$ as from $b$ to $a$, which holds in general for bipartite undirected random graphs and any interacting undirected graphs.

The network generation works as follows. First, each node independently draws its pair of intra-degree and inter-degree, $(k_a, k_b)$, from its network's degree distribution, $p_a$ or $p_b$. Such i.i.d. degree sequences (one for each network) are drawn until the sum of the degrees \emph{within} each network is even and the sum of the inter-degrees from $a$ to $b$ equals the sum of the degrees from $b$ to $a$ (since the edges are undirected). (Note that this second requirement conveniently vanishes for directed inter-edges.) Once valid degree sequences are drawn, each node's ``half-edges'' (or ``edge stubs'') of the two flavors (namely, ``toward $a$'' and ``toward $b$'') are wired randomly, as in the configuration model (potentially with correlation between degrees, which we do not consider here). 

Three features deserve attention: parallel edges, few short cycles (i.e., it is locally tree-like), and the effective inter-degree distributions. First, the configuration model permits self-loops and parallel (multiple) undirected edges between two nodes. Parallel edges can be a desirable feature for a model of cascades because it allows variable amounts of connectivity: effectively they are single edges with integer weight. For sandpile avalanches, parallel edges mean a node could shed more than one grain to a neighbor. However, for the large, sparse random graphs considered here, parallel edges are rare.

Second, we note that this configuration model generates locally tree-like graphs, which makes them amenable to branching process approximations. Many infrastructure networks are locally tree-like; electrical grids, for example, have few small loops, but they must have large loops since the electrons cannot disappear. By contrast, many social networks are emphatically \emph{not} tree-like because of the small loops caused by transitivity: ``friends of friends tend to be friends''. Such small loops and small cliques are exceedingly unlikely in tree-like graphs generated by the configuration model, since the graphs are sparse. Percolation and contact processes on clustered networks \cite{miller_percolation, newman_clustering} and ones with arbitrary distributions of small subgraphs \cite{newman_subgraphs} have recently been studied.

Finally, we calculate the \emph{effective} inter-degree distributions, which are not the input inter-degree distributions because the undirected edges between the two networks require that $a$ has as many edge stubs toward $b$ as $b$ has toward $a$. Intuitively, if the inter-degree distributions $p_{ab}(\cdot)$ and $p_{ba}(\cdot)$ have much overlap, then the effective inter-degree distributions hardly differ from the input ones. But if they profoundly differ---for example, two Poisson distributions with very different means---then the valid degree sequences, which condition on the number of inter-edges agreeing, have a very different distribution than the input degree distributions. For this example of two Poisson distributions with different means, the \emph{effective} means would be much closer to one another, since valid degree sequences occur more frequently between the two means.

\begin{lem}\label{inter_degree_lemma}
Let $\vec X, \vec Y$ be the random variables for the inter-degree sequences of $a$ and $b$, which are vectors of length $N_a, N_b >0$, respectively. We denote $\Sigma \vec x \equiv \sum_{i=1}^n x_i$ to be the sum of the entries in the vector.  Then the effective inter-degree sequence for $a$ is 
\begin{align}
\Pr(\vec X = \vec k \mid \Sigma \vec Y = \Sigma \vec k) = Pr(\vec X = \vec k) \, p_{ba}^{*N_b}\big ( \Sigma \vec k \big)
\end{align}
where $p_{ba}^{*N_b}(\cdot)$ is $p_{ba}$ convolved $N_b$ many times.
\end{lem}
\noindent \textsc{Proof. } Using the independence of $\vec X$ and $\vec Y$ and by summing over all $\vec \ell \in \mathbb{Z}^{N_b}$, we have
\begin{align}
\Pr(\vec X = \vec k \mid \Sigma \vec Y = \Sigma \vec k) &= \frac{\Pr(\vec X = \vec k,\Sigma \vec Y = \Sigma \vec k)}{\Pr(\Sigma \vec Y = \Sigma \vec k)}\notag \\
&= \frac{\sum_{\vec \ell}\Pr(\vec X = \vec k, \vec Y = \vec \ell, \Sigma \vec \ell = \Sigma \vec k)}{\Pr(\Sigma \vec Y = \Sigma \vec k)}\notag \\
&= \frac{\Pr(\vec X = \vec k) \sum_{\vec \ell} \Pr(\vec Y = \vec \ell, \Sigma \vec \ell = \Sigma \vec k \mid \vec X = \vec k)}{\Pr(\Sigma \vec Y = \Sigma \vec k)} \label{independence_step} \\
&= \Pr(\vec X = \vec k) \sum_{\vec \ell} \Pr(\vec Y = \vec \ell \mid \vec X = \vec k, \Sigma \vec Y = \Sigma \vec k)\notag \\
&= \Pr(\vec X = \vec k) \underbrace{(p_{ba} * p_{ba} * ... * p_{ba})}_\textrm{$N_b$ many convolutions} \big (\Sigma \vec k \big).\label{Nb_fold_convolution}
\end{align}
Eq. \eqref{independence_step} follows from the independence of $\vec X$ and $\vec Y$, and Eq. \eqref{Nb_fold_convolution} follows from the definition of the convolution.\hfill $\Box$ \\

Lemma \ref{inter_degree_lemma} says that the \emph{effective} inter-degree distribution of $a$ is a product of the input inter-degree distribution $p_{ab}$ of $a$ with the inter-degree distribution $p_{ba}$ of $b$ convolved $N_b$ times and evaluated at the sum of the entries of the input to $p_{ab}$. In the literature this effect is often overlooked in theoretical calculations. In practice, when generating bipartite graphs (or other interacting networks) from degree distributions using the configuration model, a common, quick solution is to draw degree sequences from their distributions, and then repeatedly choose a node uniformly at random from anywhere in the network and re-draw its degree until the degree sequences are valid. However, this method suffers from the above effect, which often is subtle but can be substantial if the degree distributions have ``little overlap''. The inter-degree distributions considered here have ``much overlap'', so the effective inter-degree distribution is approximately the input one, and the correction factor in Eq. \eqref{Nb_fold_convolution} can be neglected.

\section{Avalanche size}
Most grains of sand dropped onto the network do not topple any nodes; instead, they simply increase a node's load, but not beyond its capacity. Some grains of sand topple a node that in turn may topple a few others. Even fewer trigger large avalanches that topple nearly the entire network before equilibrium is restored. We are interested in the asymptotic distribution of the sizes of avalanches, after many grains of sand have been dropped.

\subsection{Two measures of avalanche size: topplings and sheddings}
To that end, we measure the size of avalanche in two ways: (1) the numbers $t_a, t_b$ of toppling events in $a$ and in $b$, respectively, and (2) the numbers of grains of sand that are shed from one network to itself or to the other network. For short we call the former ``$a$-topplings'' and ``$b$-topplings'', and we call the latter ``$od$-sheddings'', where $o,d \in \{a,b\}$ are the ``origin'' and ``destination'' networks of the shedded grain of sand. For example, the toppling of an $a$-node that has $k_a = 5$ neighbors in $a$ and $k_b = 1$ neighbor in $b$ counts as one $a$-toppling, five $aa$-sheddings and one $ab$-shedding. For a wide range of sandpile models, the various measures of avalanche size---size, area, perimeter, duration, maximal distance, radius of gyration, etc.---scale against each other in the form of power laws \cite{benhur}, which suggests that it suffices to study just a couple of measures.

Since the two networks have different degree distributions, the distributions of the number of topplings depends on the network in which the first grain is dropped. As an example, a grain dropped in a dense network $a$, weakly coupled to a sparse network $b$, would likely cause larger avalanches in $b$---compared to avalanches begun in $b$---since $a$ has more total capacity, so its large avalanches---the most likely ones to cross the sparse inter-network edges---overwhelm the lower-capacity network $b$. Thus we have two distributions of toppling size, with the subscript indicating the network in which the avalanche begins:
\begin{align*}
s_a(t_a, t_b) &\equiv \text{prob. a grain dropped in $a$ causes $t_a$ topplings in $a$ and $t_b$ in $b$},\\
s_b(t_a, t_b) &\equiv \text{prob. a grain dropped in $b$ causes $t_a$ topplings in $a$ and $t_b$ in $b$}.
\end{align*}
These toppling size distributions (`$s$' for ``size'') count the initial toppling, and so they satisfy
\begin{align*}
s_a(0, t_b) &= 0 \qquad \forall \,\, t_b \geq 1,\\
s_b(t_a, 0) &= 0 \qquad \forall \,\, t_a \geq 1,
\end{align*}
since, for example, an $a$-node cannot topple $b$-nodes unless it topples first. Note that $s_a$ and $s_b$ are probability distributions in the asymptotic limit, in the sense that they are frequencies of avalanche sizes after the network has undergone many cascades. For each avalanche size distribution, we define the associated generating functions
\begin{align*}
\mathcal{S}_a(\tau_a, \tau_b) &= \sum_{t_a,t_b = 0}^\infty s_a(t_a, t_b) \tau_a^{t_a} \tau_b^{t_b},\\
\mathcal{S}_b(\tau_a, \tau_b) &= \sum_{t_a,t_b = 0}^\infty s_b(t_a, t_b) \tau_a^{t_a} \tau_b^{t_b},
\end{align*}
for $\tau_a, \tau_b \in \mathbb{C}$ with $\tau_a, \tau_b \leq 1$.

For interacting networks, \emph{sheddings} rather than \emph{topplings} are a more natural way to mathematically study avalanche size, for reasons discussed below. Recall the definition of an \emph{$od$-shedding}: \begin{defin}An \emph{$od$-shedding} is the event that a grain of sand is shed from network $o \in \{a,b\}$ to network $d \in \{a,b\}$. (We use `$o$' for ``origin'', `$d$' for ``destination''.)\end{defin} In order to calculate the distributions $s_a, s_b$ of the number of \emph{topplings} in $a$ and $b$, we must first calculate the distributions of the number of \emph{sheddings} of the four origin-destination types, $aa, ab, ba, bb$. From these shedding distributions we then calculate the avalanche size distributions $s_a, s_b$.

We denote by $\rho_{aa}, \rho_{ab}, \rho_{ba}, \rho_{bb}$ the probability distributions of the number of $od$-sheddings of the four types ($aa, ab, ba, bb$) in an avalanche caused by an initial shedding of the type specified in the subscript. Like the avalanche size distributions $s_a, s_b$, these are defined asymptotically, after many grains of sand have been dropped on the network. The probability generating functions $\mathcal{P}_{od}$ associated to the shedding distributions $\rho_{od}$ (where $o,d \in \{a,b\}$) are
\begin{align*}
\mathcal{P}_{od}(\sigma_{aa}, \sigma_{ab}, \sigma_{ba}, \sigma_{bb}) &= \sum_{r_{aa}, r_{ab}, r_{ba}, r_{bb} = 0}^\infty \rho_{od}(r_{aa}, r_{ab}, r_{ba}, r_{bb}) \sigma_{aa}^{r_{aa}} \sigma_{ab}^{r_{ab}} \sigma_{ba}^{r_{ba}} \sigma_{bb}^{r_{bb}},
\end{align*}
for $\sigma_{aa}, \sigma_{ab}, \sigma_{ba}, \sigma_{bb} \in \mathbb{C}$ with absolute value $\leq 1$.

\subsection{Sheddings are topplings on the line graph}
Whereas cascades of \emph{topplings} correspond to cascades of nodes in the \emph{original graph}, cascades of \emph{sheddings}, in a sense formalized below, correspond to cascades on the nodes of the \emph{line graph} of the original graph. (Recall that the line graph \cite{line_graph} of a graph $G$ is a graph with nodes labels given by edges in $G$, and two nodes in the line graph are connected if and only if the corresponding edges in $G$ share a common vertex.)

Now we make the connection between sheddings and the line graph precise. For a given avalanche let the \emph{toppling digraph} be the directed graph in which there's an edge from node $u$ to node $v$ if $u$ shed at least one grain of sand to $v$ in the avalanche. Similarly, let the \emph{shedding digraph} be the directed graph in which two nodes labeled `$u \rightarrow v$' and `$w \rightarrow x$' are connected if $v=w$ and a shedded grain from $u$ to $v$ caused $v=w$ to topple and to shed a grain of sand to $x$---i.e., if $u$ toppled, which toppled $v=w$.

\begin{prop}If the graph is a tree, then the shedding digraph is the line graph of the toppling digraph.\end{prop}
\noindent \textsc{Proof. } If node $u$ topples, which eventually topples $v$, then every node on the unique path between $u$ and $v$ toppled in succession because the graph is a tree, and hence the successive directed edges along the path are connected in the shedding digraph. Conversely, if $u$ is not connected to $v$ in the toppling digraph, then---again since the graph is a tree---it never occurred that $v$ toppled immediately after $u$ toppled, and so for any node $y$ there does not exist an edge from $(u,v)$ to $(v,y)$ in the shedding digraph.\hfill $\Box$\\
\begin{rem}If the toppling digraph is only approximately tree-like, then this becomes less precise, but in a sense the shedding digraph is approximately the line graph of the toppling digraph. There may be a way to encode the avalanche in a weighted digraph that preserves all the information about ``who toppled whom'' in the avalanche.\end{rem}

\subsection{Branching process approximations}
Cascades in networks can be approximated by a multiplicative branching process if there are few loops of small or intermediate size. Said differently, a growing avalanche is approximately a tree with branches that grow and terminate independently. This approximation works well for locally tree-like networks, such as the random graphs generated here using the configuration model; since the edges are sparse and wired uniformly at random, the chance of closing a triangle (or any other small loop) is negligibly small. Two loosely coupled networks appear to be approximately tree-like as long as the coupling is sparse, as verified for coupled Erd\H{o}s-R\'{e}nyi and power-law graphs in \cite{leichtdsouza}. For two coupled networks, the avalanches can be approximated by two-type and four-type Galton-Watson processes for the distributions of topplings and sheddings, respectively, which we formalize next.

The fundamental tool in branching processes is the ``children'' or ``branch'' distribution, the distribution of the number of ``children'' objects of the various types generated by a ``parent'' object of a certain type. Many probabilistic properties of the tree generated by an initial population are calculated using the branch distribution (or even just its moments). It turns out that it is more natural to write down the branch distribution of sheddings rather than topplings.

\subsection{Shedding branch distributions $q_{aa}, q_{ab}, q_{ba}, q_{bb}$}\label{Subsection:shedding_branch}
To calculate the shedding distributions $\rho_{aa}, \rho_{ab}, \rho_{ba}, \rho_{bb}$, we first must determine the distributions $q_{aa}, q_{ab}, q_{ba}, q_{bb}$ of the number of ``children sheddings'' caused by a ``parent shedding'' of the four types ($aa, ab, ba ,bb$) specified in the subscript.

\begin{defin}The \emph{branch distributions} $q_{od}(r_{da},r_{db})$, where $od \in \{aa, ab, ba, bb\}$, are the probabilities that a grain of sand shedded from an ``origin'' node in network $o \in \{a,b\}$ to a ``destination'' node in network $d \in \{a,b\}$ topples the destination node, which in turn sheds one grain to each of its $r_{da}$ many $a$-neighbors and $r_{db}$ many $b$-neighbors.\end{defin}
Said differently, $q_{od}(r_{da},r_{db})$ is the chance that an $od$-shedding causes $r_{da}$ many $da$-sheddings and $r_{db}$ many $db$-sheddings. In the language of Galton-Watson processes, $q_{od}(r_{da}, r_{db})$ is the distribution of the number of ``children sheddings'' of types $da, db \in \{aa, ab, ba, bb\}$ caused by a ``parent'' shedding of type $od \in \{aa, ab, ba, bb\}$. The ``children'' sheddings necessarily originate in network $d \in \{a,b\}$.

To determine the $q_{od}$, consider a grain of sand that has just been shedded from an ``origin node'' to a ``destination node''. First we estimate the chance that this grain of sand topples the destination node (the node to which it is shed). Empirically, similar to \cite{goh_sandpile_powerlaw} we find that the amounts of sand on nodes are approximately uniformly distributed from zero to one less their degree asymptotically---there is no typical amount of sand---so the probability that the destination node topples is the probability that it has one fewer grain of sand than its total degree: $1 / (k_a + k_b)$. (Later we mention correction terms that account for toppling due to receiving multiple grains of sand at once, which is possible only if there are loops in the network, and so they can be neglected for the locally tree-like networks considered here.)

We are interested not only in whether the destination node topples but also in how many grains it sheds to each network, which is the destination node's intra- and inter-degree. The asymmetry of the coupling between networks implies that the degree distribution of the destination node depends on the network to which the origin node belongs. For example, suppose the inter-edges only connect intra-hubs (the nodes with high intra-degree); then a grain traveling an inter-edge is more likely than one traveling an intra-edge to land on a hub (even though hubs are already likely to be found at the end of intra-edges!). This is why determining the branch distributions $q_{od}$ of sheddings is easier than determining the branch distributions of topplings: we must carry two indices in the subscript to use the fact that the distribution of the degree of the destination node in network depends on the network to which the origin node belongs.

If the networks $a$ and $b$ are connected by sparse edges \emph{without intra-degree correlations}, then---to good approximation---a grain traveling from, say, network $a$ to network $b$ arrives at an inter-edge stub in $b$ chosen uniformly at random. Similarly, if the intra-edge stubs are joined without degree correlations, then a grain traveling an intra-edge is equally likely to arrive at any other intra-edge stub. Since a $d$-node with $r_{do} \geq 1$ many $o$-neighbors is $r_{do}$ times more likely than a $d$-node with one $o$-neighbor to receive a grain of sand, for $o, d \in \{a,b\}$ we have
\begin{align}
q_{od}(r_{da}, r_{db}) &= \frac { r_{do} p_d(r_{da}, r_{db})} {\langle k_{do} \rangle } \frac{1}{r_{da} + r_{db}} \,\,\,\,\,\,\,\,\,\,\,\, \text{for } r_{da} + r_{db} > 0,\label{qab}
\end{align}
and we set
\begin{align}
q_{od}(0,0) &:= 1- \sum_{r_{da} + r_{db} > 0} q_{od}(r_{da}, r_{db}).\label{qab_normalization}
\end{align}
The denominator $\langle k_{do} \rangle := \partial_{\omega_o}G_d(1,1)$ is the expected number of edges from $d$ to $o$; it normalizes the numerator $r_{do} p_d(r_{da}, r_{db})$ to be a uniform probability distribution over the edge stubs from network $d$ to network $o$. Eq. \eqref{qab_normalization} is the probability that the destination node does not topple---i.e., that it has fewer grains than one less than its total degree.

The justification of Eq. \eqref{qab} is as follows. Define the following four random variables for a node in network $b$:
\begin{enumerate}
\item{$K_a \in \mathbb{Z}_{\geq 0}$, the node's $a$-degree}
\item{$K_b \in \mathbb{Z}_{\geq 0}$,  the node's $b$-degree}
\item{$G \in \mathbb{Z}_{\geq 0}$, the number of grains on it}
\item{$R \in \{a, b\}$, whether it receives a grain from network $a$ or from $b$}
\end{enumerate}
Next we define, for network $b$, the probability distribution $\Pr(K_a, K_b, G, R)$ on $\mathbb{Z}_{\geq 0} \times \mathbb{Z}_{\geq 0} \times \mathbb{Z}_{\geq 0} \times \{a, b\}$. Then the chance that a $b$-node receives a grain from an $a$-node and topples its grains to $k_a$ $a$-neighbors and $k_b$ $b$-neighbors is
\begin{align*}
q_{ab}(k_a, k_b) &= \Pr(K_a = k_a, K_b = k_b, G = k_a + k_b - 1, R = a)\\
&= \Pr(K_a = k_a, K_b = k_b) \cdot \Pr(G = k_a + k_b - 1\mid K_a = k_a, K_b = k_b) 
\\ & \,\,\,\,\,\,\,\,\,\,\,\,\,\,\,\,\,\,\,\,\,\, \cdot \Pr(R = a \mid K_a = k_a, K_b = k_b, G = k_a + k_b - 1) \\
&= p_b(k_a, k_b) \cdot \frac {1} {k_a + k_b} \cdot \frac {k_a} {\sum_{i = 1}^\infty \sum_{j = 0}^\infty i p_b(i, j)}
\end{align*}

\begin{rem}If the degree distributions of networks $a$ and $b$ are independent, meaning they can be written as products of intra- and inter-degrees
\begin{align*}
p_a(k_a, k_b) &= p_{aa}(k_a) p_{ab}(k_b),\\
p_b(k_a, k_b) &= p_{ba}(k_a) p_{bb}(k_b),
\end{align*}
then the branch distributions simplify to
\begin{align*}
q_{od}(r_{da}, r_{db}) = \frac { r_{do} p_{da}(r_{da}) p_{db}(r_{db})} { \langle k_{do} \rangle} \frac{1}{r_{da} + r_{db}}
\end{align*}
where (as above) we denote averages by $\langle \cdot \rangle$ (e.g., $\langle k_{aa} \rangle = \sum_{i=1}^\infty i p_{aa}(i))$). As above, these equations hold for $k_a, k_b$ not both zero; at the origin they equal one minus their values everywhere else in $\mathbb{Z}_{\geq 0}^2$.\end{rem}

Note that the branch distributions $q_{od}$ are the distributions of sheddings \emph{at the next time step}, so they depend on only two inputs---$k_{da}$ and $k_{db}$---because a $d$-toppling can only cause $da$- and $db$-sheddings. By contrast, the shedding distributions $\rho_{od}$ count the \emph{total} number of $aa$-, $ab$-, $ba$- and $bb$-sheddings in avalanches initiated by an $od$-shedding. Since an $od$-shedding could lead to sheddings of any type in the ensuing avalanche, the $\rho_{od}$ depend on four inputs: $r_{aa}, r_{ab}, r_{ba}, r_{bb}$.

If the inter-edges are assortative, disassortative, or some other pattern deviating from independence, then the expressions for $q_{ab}$ and $q_{ba}$ need to change. For example, if the edges between the networks are assortative and the node that's shedding has high intra-degree and is in $a$, then $q_{ab}$ should be larger for nodes with high intra-degree. This needs a new, more general framework than the $q_{od}$ above, since we need to take into account the intra-degrees of the two nodes, which do not appear anywhere in our equations for the $q$'s. Similarly, if $a$ and $b$ have intra-assortativity (or some other correlation), then we need to modify $q_{aa}$ and $q_{bb}$ to account for these correlations.

The $q_{od}$ are only approximations to the avalanche dynamics, and one could add correction terms for greater accuracy. First, if the network is not approximately tree-like (such as the Watts-Strogatz model on 1D lattices or random graphs with triangles added), one could add a correction for loops by computing the chance that a node is two grains away from toppling and computing all the ways of receiving two grains of sand in a given time step.

A second source of correction terms is the dissipation rate of sand $f$. To shed $k_a, k_b$ grains, one must consider all the ways of shedding at least that many, since arbitrarily many grains could vanish. These corrections to the $q_{od}$ are binomially distributed with parameter $f$:
\begin{align*}
\tilde q_{od}(r_{da}, r_{db}) = \sum_{k_a=r_{da}}^\infty \sum_{k_b=r_{db}}^\infty q_{od}(k_a, k_b) (1-f)^{r_{da}+r_{db}} f^{k_a+k_b-r_{da}-r_{db}} \binom{k_a}{r_{da}} \binom{k_b}{r_{db}},
\end{align*}
since $1-f$ is the chance that a grain does not vanish, and $r_{da}$ of the grains shed to $a$ and $r_{db}$ of the grains shed to $b$ must survive. Ideally, the dissipation rate is strong enough that the system does not become overloaded with sand, but not too strong that it plays a significant role in the avalanche sizes compared to the role that $q(0,0)$ plays in inhibiting avalanches. Thus, as a rule of thumb one should choose $f$ to be one or two orders of magnitude smaller than $\min\{q_{od}(0,0) | o,d \in \{a,b\}\}$.

Finally, we define the generating functions $\{\mathcal{Q}_{od} : od \in \{aa, ab, ba, bb\}\}$ associated to the four branching distributions $\{q_{od}| od \in \{aa, ab, ba, bb\}\}$:
\begin{align*}
\mathcal{Q}_{od}(\sigma_{da}, \sigma_{db}) &= \sum_{r_{da}, r_{db} = 0}^\infty q_{od}(r_{da}, r_{db}) \sigma_{da}^{r_{da}} \sigma_{db} ^{r_{db}} \qquad \text{for } \sigma_{da}, \sigma_{db} \in \mathbb{C}.
\end{align*}

\subsection{Shedding self-consistency equations}
Next we connect the generating functions for the total shedding size $\mathcal{P}_{od}$ to the branch distribution $\mathcal{Q}_{od}$ in so-called ``self-consistency equations''. First, we denote
\begin{align*}
\vec \sigma &:= (\sigma_{aa}, \sigma_{ab}, \sigma_{ba}, \sigma_{bb}),\\
\vec{\mathcal{P}}(\vec \sigma) &:= (\mathcal{P}_{aa}(\vec \sigma), \mathcal{P}_{ab}(\vec \sigma), \mathcal{P}_{ba}(\vec \sigma), \mathcal{P}_{bb}(\vec \sigma)),\\
\vec{\mathcal{Q}}(\vec \sigma) &:= (\mathcal{Q}_{aa}(\vec \sigma), \mathcal{Q}_{ab}(\vec \sigma), \mathcal{Q}_{ba}(\vec \sigma), \mathcal{Q}_{bb}(\vec \sigma)),\\
\mathcal{Q}_{od}(\vec{\mathcal{P}}(\vec \sigma)) &:= \mathcal{Q}_{od}(\mathcal{P}_{da}(\vec \sigma), \mathcal{P}_{db}(\vec \sigma)).
\end{align*}
By the theory of multi-type, multiplicative branching (Galton-Watson) processes (e.g., Eq. (10.3) of \cite{harris}), we have
\begin{align}
\vec{\mathcal{P}}(\vec \sigma) = \vec \sigma \cdot \vec{\mathcal{Q}}(\vec{\mathcal{P}}(\vec \sigma)),\label{branching_self_consistency}
\end{align}
where $\cdot$ is the usual dot product. Written more explicitly,
\begin{align}
\mathcal{P}_{aa} &= \sigma_{aa} \mathcal{Q}_{aa}(\mathcal{P}_{aa}, \mathcal{P}_{ab}), \label{Paa}\\
\mathcal{P}_{ab} &= \sigma_{ab} \mathcal{Q}_{ab}(\mathcal{P}_{ba}, \mathcal{P}_{bb}), \label{Pab}\\
\mathcal{P}_{ba} &= \sigma_{ba} \mathcal{Q}_{ba}(\mathcal{P}_{aa}, \mathcal{P}_{ab}), \label{Pba}\\
\mathcal{P}_{bb} &= \sigma_{bb} \mathcal{Q}_{bb}(\mathcal{P}_{ba}, \mathcal{P}_{bb}), \label{Pbb}
\end{align}
where all the $\mathcal{P}$'s are evaluated at $\vec \sigma \equiv (\sigma_{aa}, \sigma_{ab}, \sigma_{ba}, \sigma_{bb})$. In general,
\begin{align*}
\mathcal{P}_{od}(\vec \sigma) = \sigma_{od} \mathcal{Q}_{od}(\mathcal{P}_{da}, \mathcal{P}_{db}).
\end{align*}
In words, Eq. \eqref{Paa}, for example, says that to get an avalanche of sheddings of the four types ($aa, ab, ba, ba$) starting from an $aa$-shedding, the cascade tree begins with an $aa$-shedding (that's the $\sigma_{aa}$ out front), which causes at the next time step a number of other $aa$- and $ab$-sheddings distributed according to $\mathcal{Q}_{aa}$, each of which in turn cause numbers of sheddings distributed according to $\mathcal{P}_{aa}$ and $\mathcal{P}_{ab}$. 

\subsection{Two null hypotheses for interacting networks}
To fully understand the effects of interactions between distinct networks (or of modular structure within a network) we need to compare the systems described above to an appropriate null-model.  Below are two different approaches. The first neglects node flavor when calculating total avalanche sizes, while the second assumes that the networks are uncoupled. 


\begin{nullhypothesis}\label{flavorlessview}\emph{[Flavorless View]} Avalanches in interacting networks should be considered in the ``flavorless view,'' which does not distinguish what portion of the avalanche lies in each network.\end{nullhypothesis}
The distribution of total avalanche size in the ``flavorless view'' can be recovered via convolution:
\begin{align}
s_a^*(t) &= \sum_{i=0}^t s_a(i, t-i),\label{s_a_total}\\
s_b^*(t) &= \sum_{i=0}^t s_b(i, t-i).\label{s_b_total}
\end{align}
Note that here the total avalanche size still depends on in which network the first grain is dropped; if it is dropped uniformly at random, then the distribution of avalanche sizes is the convex combination of \eqref{s_a_total}, \eqref{s_b_total} with weights equal to the proportions of $a$- and $b$-nodes.

The flavorless view of interacting networks is another way of viewing the system; by contrast, the ``uncoupled null hypothesis'' supposes that the networks are not connected at all.
\begin{nullhypothesis}\label{uncoupled}\emph{[Uncoupled]} The two networks are uncoupled (i.e., not connected by any edges).\end{nullhypothesis}Avalanches on uncoupled networks are independent, so the avalanche size distribution on uncoupled networks is the product measure of the avalanche size distributions on each network.

Another null hypothesis for interacting networks, which we do not use here, considers the properties of a single network with equivalent size and connectivity (i.e., degree distribution) as the system of modular or interdependent networks \cite{leichtdsouza}.


\subsection{Shedding equations in the flavorless view}
The branching process framework for the flavorless view is straightforward to obtain. Let $\rho_a, \rho_b$ denote the fraction of $a$- and $b$-nodes (i.e., $\rho_a = \frac{N_a} {N_a + N_b}, \rho_b = \frac{N_b} {N_a + N_b}$). For $o, d \in \{a,b\}$ let
\begin{align*}
q_{od}^*(k) := \sum_{i=0}^k q_{od}(i, k-i)
\end{align*}
denote the sum of $q_{od}$ along the diagonal $\{(i, k-i): i \in \{0, 1, ..., k\}\}$.

If grains of sand are dropped uniformly at random, then the chance that it lands on an $a$-node is $\rho_a$ (and similarly $\rho_b$ for $b$-nodes). Thus the avalanche size distribution in the flavorless view of the two coupled networks is
\begin{align*}
s(t) &= \rho_a s_a^*(t) + \rho_b s_b^*(t) \qquad \forall \,\,t \geq 0,
\end{align*}
where $s_a^*, s_b^*$ are the distributions of the total sizes of avalanches begun in $a, b$, defined in Eqs. \eqref{s_a_total}, \eqref{s_b_total}.

Consider a grain of sand traveling along an edge; if the graph is fully connected, then the flavors of the grain's origin and destination nodes are approximately independent, so we can approximate the branching distribution in the flavorless view as
\begin{align*}
q(k) &= \rho_a^2 q_{aa}^*(k) + \rho_a \rho_b \big( q_{ab}^*(k) + q_{ba}^*(k) \big)+ \rho_b^2 q_{bb}^*(k) \qquad \forall \,\, k \geq 0.
\end{align*}
These distributions generate
\begin{align*}
\mathcal{P}(\omega) = \sum_{t=0}^\infty s(t) \omega^t, \qquad \mathcal{Q}(\omega) = \sum_{k=0}^\infty q(k) \omega^k,
\end{align*}
which are related according to the self-consistency equation
\begin{align*}
\mathcal{P}(\omega) = \mathcal{Q}(\mathcal{P}(\omega)).
\end{align*}

\subsection{Toppling branch distributions $u_a, u_b$}
Although it is more natural to write down the branch distribution $q_{od}$ of sheddings rather than that of topplings, and although the numbers of sheddings tell a different picture about the cascade (the total amount of sand shed rather than the numbers of topplings), it is convenient to reduce the dimension of avalanche statistics by measuring how many nodes in each network toppled in an avalanche rather than how many grains were shed from one network to another. As a result, we next derive the corresponding toppling branch distributions $u_a, u_b$ from the shedding branch distributions $q_{od}$, and then we solve for the toppling distributions $s_a, s_b$.

The key insight is that a node topples if and only if it sheds at least one grain of sand. Thus a grain traveling from a network $o$ to network $d$ topples its destination node with probability $1-q_{od}(0,0)$. If an $a$-node topples, what is the chance that it topples $t_a$ more $a$-nodes and $t_b$ more $b$-nodes in the next time step? Denoting this branch distribution by $u_a(t_a, t_b)$, we have
\begin{align*}
u_a(t_a, t_b) = \sum_{k_a = t_a}^\infty \sum_{k_b = t_b}^\infty p_a(k_a, k_b) &(1-q_{aa}(0,0))^{t_a} q_{aa}(0,0)^{k_a - t_a} \binom{k_a}{t_a} \times \\&\times  (1-q_{ab}(0,0))^{t_b} q_{ab}(0,0)^{k_b - t_b} \binom{k_b}{t_b} \notag
\end{align*}
since the node must have at least $t_a$ many $a$-neighbors and at least $t_b$ many $b$-neighbors, only $t_a, t_b$ of which topple---which are binomially distributed. Similarly, the branch distribution $u_b(t_a, t_b)$ is
\begin{align*}
u_b(t_a, t_b) = \sum_{k_a = t_a}^\infty \sum_{k_b = t_b}^\infty p_b(k_a, k_b) & (1-q_{ba}(0,0))^{t_a} q_{ba}(0,0)^{k_a - t_a} \binom{k_a}{t_a} \times \\ & \times (1-q_{bb}(0,0))^{t_b} q_{bb}(0,0)^{k_b - t_b} \binom{k_b}{t_b}. \notag
\end{align*}
Denoting the associated generating functions by $\mathcal{U}_a(\tau_a, \tau_b), \mathcal{U}_b(\tau_a, \tau_b)$, we obtain the self-consistency relations
\begin{align}
\mathcal{S}_a &= \tau_a \, \mathcal{U}_a (\mathcal{S}_a, \mathcal{S}_b),\label{Ua}\\
\mathcal{S}_b &= \tau_b \, \mathcal{U}_b (\mathcal{S}_a, \mathcal{S}_b),\label{Ub}
\end{align}
where each $\mathcal{S}$ is evaluated at $(\tau_a, \tau_b)$.

\subsection{Summary of the distributions}
Before solving the self-consistency equations, we pause to summarize the branching process framework's distributions and generating functions. (Recall that another common name for ``branch distribution'' is ``children distribution''.)
 \begin{table}[h]
\caption{Summary of the distributions and their generating functions} 
\centering
  \begin{tabular}{ c | c  c }
    & distribution & generating function \\ \hline
    degree & $p_a(k_a,k_b), p_b(k_a,k_b)$ & $G_a(\omega_a, \omega_b), G_b(\omega_a, \omega_b)$  \\ \hline
    toppling size & $s_a(t_a,t_b), s_b(t_a,t_b)$ & $\mathcal{S}_a(\tau_a, \tau_b), \mathcal{S}_b(\tau_a, \tau_b)$  \\ \hline
    toppling branch & $u_a(t_a, t_b), u_b(t_a, t_b)$ & $\mathcal{U}_a(\tau_a, \tau_b), \mathcal{U}_b(\tau_a, \tau_b)$  \\ \hline
    shedding size & $\rho_{od}(r_{aa}, r_{ab}, r_{ba}, r_{bb})$ & $\mathcal{P}_{od}(\sigma_{aa}, \sigma_{ab}, \sigma_{ba}, \sigma_{bb})$  \\ \hline
    shedding branch & $q_{od}(r_{da},r_{db})$ & $\mathcal{Q}_{od}(\sigma_{da},\sigma_{db})$  \\ \hline
  \end{tabular}
\label{tab:distributions}
\end{table}

\subsection{Solving the self-consistency equations}
We wish to solve Eqs. \eqref{Paa}, \eqref{Pab}, \eqref{Pba}, \eqref{Pbb} for $\mathcal{P}_{aa}, \mathcal{P}_{ab}, \mathcal{P}_{ba}, \mathcal{P}_{bb}$, and Eqs. \eqref{Ua}, \eqref{Ub} for $\mathcal{U}_{a}, \mathcal{U}_{b}$. Then we obtain their underlying probability distributions by reading the coefficients or by differentiating, since, for example,
\begin{align*}
s_a(t_a, t_b) = \frac{1}{t_a! t_b!} \frac{\partial^{t_a}}{\partial \tau_a^{t_a}} \frac{\partial^{t_b}}{\partial \tau_b^{t_b}} \mathcal{S}_a(\tau_a, \tau_b) \bigg|_{\tau_a=0, \tau_b=0}.
\end{align*}
As described in \cite{watts_newman_strogatz}, since numerical differentiation is prone to machine-precision errors, to obtain the best precision it is preferred to use Cauchy's integration formula,
\begin{align*}
s_a(t_a, t_b) = \frac{1}{(2 \pi i)^2} \int \hspace{-2mm} \int_D \, \! \frac{\mathcal{S}_a(\tau_a, \tau_b)}{\tau_a^{t_a+1}\tau_b^{t_b+1}} \, d\tau_a d\tau_b,
\end{align*}
integrating over a domain $D \subset \mathbb{C}^2$ that is the Cartesian product of the largest contours that contain the origin and no poles of the generating function. (The generalization of Cauchy's integration formula to multiple variables can be found in, for example, Theorem 2.1.1 of \cite{complexanalysis_multivariable}.)

In practice, the self-consistency equations \eqref{Paa}-\eqref{Pbb}, \eqref{Ua}-\eqref{Ub} are transcendental and difficult to invert. However, a generalization of Lagrange's expansion to several variables due to I. J. Good \cite{good_lagrange} provides general tools for inverting self-consistency equations of multi-type branching processes. For example, Theorem 9 of \cite{good_lagrange} allows us to compute coefficients of the distributions, one at a time. We state the theorem here for the shedding distributions, but it holds for any multi-type branching processes in which each type has a positive chance of giving birth to zero children.

\begin{theo}[Good 1960] Suppose that $\vec{\mathcal{Q}}(\vec \sigma)$ is analytic in a neighborhood of the origin and that $\mathcal{Q}_{od}(\vec 0) \equiv q_{od}(0,0) \not = 0$ for all $o,d \in \{a,b\}$. The probability that the whole avalanche consists of exactly $m_{od}$ many $od$-sheddings, starting from $i_{od}$ sheddings, is the coefficient of
\begin{align*}
\sigma_{aa}^{m_{aa} - i_{aa}} \sigma_{ab}^{m_{ab} - i_{ab}} \sigma_{ba}^{m_{ba} - i_{ba}} \sigma_{bb}^{m_{bb} - i_{bb}} 
\end{align*} 
in
\begin{align}
\mathcal{Q}_{aa}^{m_{aa}} \mathcal{Q}_{ab}^{m_{ab}} \mathcal{Q}_{ba}^{m_{ba}} \mathcal{Q}_{bb}^{m_{bb}} \bigg | \bigg | \delta_\mu^\nu - \frac {\sigma_\mu}{\mathcal{Q}_\mu} \frac {\partial \mathcal{Q}_\mu}{\partial \sigma_{\mu}} \bigg | \bigg |\label{determinant_equation}
\end{align}
where $\delta_\mu^\nu$ is the Kronecker delta, $|| \cdot ||$ is the determinant, and $\mu, \nu$ run over $\{aa, ab, ba, bb\}$. \label{theorem9}\end{theo}

Even more useful is Theorem 10 of \cite{good_lagrange}, which explicitly gives the generating function solutions rather than just one coefficient at a time. Since we use this theorem to compute the toppling size distributions, we state it for the toppling size generating functions. Denote $\vec \tau \equiv (\tau_a, \tau_b)$, $\vec{\mathcal{S}} \equiv (\mathcal{S}_a, \mathcal{S}_b)$ and let $h(\vec \tau) \equiv \tau_a^{r_1} \tau_b^{r_2}$ for arbitrary nonnegative integers $r_1, r_2$.
\begin{theo}[Good 1960] Under the same conditions as in Theorem \ref{theorem9} (namely, $\vec{\mathcal{U}}(\vec \tau)$ is analytic in a neighborhood of the origin and $\mathcal{U}_a(\vec 0) \not = 0, \mathcal{U}_b(\vec 0) \not = 0$), we have
\begin{align}
h(\vec{\mathcal{S}}(\vec \tau)) = \sum_{m_a, m_b = 0}^\infty  \frac{\tau_a^{m_a} \tau_b^{m_b}}{m_a! m_b!} \bigg [ \frac{\partial^{m_a+m_b}}{\partial \kappa_a^{m_a} \partial \kappa_b^{m_b}}  \bigg \{ h(\vec \kappa) \mathcal{U}_a(\vec \kappa)^{m_a}  \mathcal{U}_b(\vec \kappa)^{m_b}             \bigg | \bigg | \delta_\mu^\nu - \frac {\kappa_\mu}{\mathcal{U}_\mu} \frac {\partial \mathcal{U}_\mu}{\partial \kappa_{\mu}} \bigg | \bigg |            \bigg \} \bigg ]_{\vec \kappa = 0},\label{eq:theorem10}
\end{align}
where, as above, $\mu, \nu$ run over the types $\{a,b\}$, $\delta_\mu^\nu$ is the Kronecker delta, and $|| \cdot ||$ is the determinant.\label{theorem10}\end{theo}
This result holds for any number of types as long as each type has a positive probability of being barren; the generalization of \eqref{eq:theorem10} to four types (in order to obtain the shedding generating functions $\mathcal{P}_{od}$) is straightforward. Taking $r_1 =1, r_2=0$ in \eqref{eq:theorem10} gives $\mathcal{S}_a$, while taking $r_1=0, r_2=1$ gives $\mathcal{S}_b$. In practice, thousands of terms can be obtained by truncating the sum in \eqref{eq:theorem10} using computer algebra systems.

\section{Examples}
Next we turn to examples, beginning with the easiest (two regular graphs with one-to-one coupling) and progressively adding complexity (Bernoulli coupling, other degree distributions). For each example we give the generating functions, and in the next section we compare the theoretical predictions of the generating functions obtained from Theorem \ref{theorem10} with numerical simulation of sandpiles.

\subsection{Regular($z_a$)-One-to-One-Regular($z_b$)}
Arguably the simplest nontrivial intra-degree distribution is the delta function, which yields a random regular graph. (Approximating electrical grids by regular graphs is not unreasonable, since electrical grids in the United States have been found to have narrow degree distribution, namely exponential~\cite{amaralPNAS2000,albertPRE2004,Hines2010}). Similarly, the simplest coupling is ``one-to-one'': each node node in $a$ is connected to exactly one node in $b$, chosen uniformly at random. Note that one-to-one coupling requires that the number of $a$-nodes must equal the number of $b$-nodes. This coupling between networks is hardly realistic for, say, interdependent infrastructure, but the degree distributions are so simple---they are products of delta functions---that computing the generating functions is easy. Thus one-to-one couplings is a natural first case to solve (e.g., \cite{buldyrev}), but any conclusions for infrastructure or other real systems requires more flexible coupling.

The degree distributions are
\begin{align*}
p_a(k_a, k_b) = \delta_{z_a}(k_a) \delta_1(k_b), \qquad p_b(k_a, k_b) = \delta_{1}(k_a) \delta_{z_b}(k_b),
\end{align*}
which generate
\begin{align*}
G_a(\omega_a, \omega_b) = \omega_a^{z_a} \omega_b, \qquad G_b(\omega_a, \omega_b) = \omega_a \omega_b^{z_b}.
\end{align*}
The branch distributions simplify to
\begin{align*}
q_{aa}(r_{aa}, r_{ab}) &= q_{ba}(r_{aa}, r_{ab}) = \frac {1} {z_a+1} \delta_{z_a}(r_{aa}) \delta_1(r_{ab}),\\
q_{ab}(r_{ba}, r_{bb}) &= q_{bb}(r_{ba}, r_{bb}) = \frac {1} {z_b+1} \delta_{z_b}(r_{bb}) \delta_1(r_{ba})
\end{align*}
away from the origin, and
\begin{align*}
q_{aa}(0,0) &= q_{ba}(0,0) = \frac {z_a} {z_a + 1},\\
q_{ab}(0,0) &= q_{bb}(0,0) = \frac {z_b} {z_b + 1}
\end{align*}
at the origin. The generating functions of the branch distributions are
\begin{align*}
\mathcal{Q}_{aa}(\sigma_{aa}, \sigma_{ab}) &= \mathcal{Q}_{ba}(\sigma_{aa}, \sigma_{ab}) =  \frac {z_a}{z_a + 1} + \frac {1}{z_a + 1} \sigma_{aa}^{z_a} \sigma_{ab},\\
\mathcal{Q}_{ab}(\sigma_{ba}, \sigma_{bb}) &= \mathcal{Q}_{bb}(\sigma_{ba}, \sigma_{bb}) = \frac {z_b}{z_b + 1} + \frac {1}{z_b + 1} \sigma_{ba} \sigma_{bb}^{z_b},
\end{align*}
The self-consistency relations \eqref{branching_self_consistency} for the generating functions of the shedding distributions $\rho_{aa}, \rho_{ab}, \rho_{ba}, \rho_{bb}$ are
\begin{align*}
\mathcal{P}_{aa}(\vec \sigma) &= \sigma_{aa} \bigg [ \frac {z_a}{z_a + 1} + \frac {1}{z_a + 1} \mathcal{P}_{aa}(\vec \sigma)^{z_a} \mathcal{P}_{ab}(\vec \sigma) \bigg ],\\
\mathcal{P}_{ab}(\vec \sigma) &= \sigma_{ab} \bigg [ \frac {z_b}{z_b + 1} + \frac {1}{z_b + 1} \mathcal{P}_{ba}(\vec \sigma) \mathcal{P}_{bb}(\vec \sigma)^{z_b} \bigg ],\\
\mathcal{P}_{ba}(\vec \sigma) &= \sigma_{ba} \bigg [ \frac {z_a}{z_a + 1} + \frac {1}{z_a + 1} \mathcal{P}_{aa}(\vec \sigma)^{z_a} \mathcal{P}_{ab}(\vec \sigma) \bigg ],\\
\mathcal{P}_{bb}(\vec \sigma) &= \sigma_{bb} \bigg [ \frac {z_b}{z_b + 1} + \frac {1}{z_b + 1} \mathcal{P}_{ba}(\vec \sigma)\mathcal{P}_{bb}(\vec \sigma)^{z_b} \bigg ].
\end{align*}

By Theorem \ref{theorem9}, the probability that the whole shedding tree consists of exactly $m_{od}$ many $od$-sheddings, starting from $i_{od}$ many $od$-sheddings (for each $o,d \in \{a,b\}$), is the coefficient of
\begin{align*}
\sigma_{aa}^{m_{aa} - i_{aa}} \sigma_{ab}^{m_{ab} - i_{ab}} \sigma_{ba}^{m_{ba} - i_{ba}} \sigma_{bb}^{m_{bb} - i_{bb}} 
\end{align*} 
in
\begin{align}
\mathcal{Q}_{aa}^{m_{aa}} \mathcal{Q}_{ab}^{m_{ab}} \mathcal{Q}_{ba}^{m_{ba}} \mathcal{Q}_{bb}^{m_{bb}} \bigg | \bigg | \delta_\mu^\nu - \frac {\sigma_\mu}{\mathcal{Q}_\mu} \frac {\partial \mathcal{Q}_\mu}{\partial \sigma_{\mu}} \bigg | \bigg |\label{determinant_equation_regreg}
\end{align}
where $\delta_\mu^\nu$ is the Kronecker delta, $|| \cdot ||$ is the determinant, and $\mu, \nu$ run over $\{aa, ab, ba, bb\}$. For two regular graphs with one-to-one coupling, the matrix $\delta_\mu^\nu - \frac {\sigma_\mu}{\mathcal{Q}_\mu} \frac {\partial \mathcal{Q}_\mu}{\partial \sigma_{\mu}}$ is
\begin{align*}
\left(
\begin{array}{cccc}
 \frac{z_a+\sigma _{aa}^{z_a} \sigma _{ab}-z_a \sigma _{aa}^{z_a} \sigma _{ab}}{z_a+\sigma _{aa}^{z_a} \sigma _{ab}} & -\frac{\sigma _{aa}^{1+z_a}}{z_a+\sigma _{aa}^{z_a} \sigma _{ab}} & 0 & 0 \\
 0 & 1 & -\frac{\sigma _{ab} \sigma _{bb}^{z_b}}{z_b+\sigma _{ba} \sigma _{bb}^{z_b}} & -\frac{z_b \sigma _{ab} \sigma _{ba} \sigma _{bb}^{-1+z_b}}{z_b+\sigma _{ba} \sigma _{bb}^{z_b}} \\
 -\frac{z_a \sigma _{aa}^{-1+z_a} \sigma _{ab} \sigma _{ba}}{z_a+\sigma _{aa}^{z_a} \sigma _{ab}} & -\frac{\sigma _{aa}^{z_a} \sigma _{ba}}{z_a+\sigma _{aa}^{z_a} \sigma _{ab}} & 1 & 0 \\
 0 & 0 & -\frac{\sigma _{bb}^{1+z_b}}{z_b+\sigma _{ba} \sigma _{bb}^{z_b}} & \frac{z_b+\sigma _{ba} \sigma _{bb}^{z_b}-z_b \sigma _{ba} \sigma _{bb}^{z_b}}{z_b+\sigma _{ba} \sigma _{bb}^{z_b}}
\end{array}
\right),
\end{align*}
which has determinant
\begin{align*}
\frac{-z_b \sigma _{aa}^{z_a} \sigma _{ab} \left(-1+\sigma _{ba} \sigma _{bb}^{z_b}\right)+z_a \left(-1+\sigma _{aa}^{z_a} \sigma _{ab}\right) \left(-z_b+\left(-1+z_b\right) \sigma _{ba} \sigma _{bb}^{z_b}\right)}{\left(z_a+\sigma _{aa}^{z_a} \sigma _{ab}\right) \left(z_b+\sigma _{ba} \sigma _{bb}^{z_b}\right)}.
\end{align*}
The product
\begin{align*}
\mathcal{Q}_{aa}^{m_{aa}} \mathcal{Q}_{ab}^{m_{ab}} \mathcal{Q}_{ba}^{m_{ba}} \mathcal{Q}_{bb}^{m_{bb}} = \frac{-z_b \sigma _{aa}^{z_a} \sigma _{ab} \left(-1+\sigma _{ba} \sigma _{bb}^{z_b}\right)+z_a \left(-1+\sigma _{aa}^{z_a} \sigma _{ab}\right) \left(-z_b+\left(-1+z_b\right) \sigma _{ba} \sigma _{bb}^{z_b}\right)}{\left(z_a+\sigma _{aa}^{z_a} \sigma _{ab}\right) \left(z_b+\sigma _{ba} \sigma _{bb}^{z_b}\right)}.
\end{align*}
Thus Expression \eqref{determinant_equation_regreg} is
\begin{align*}
\frac{1}{\left(1+z_a\right) \left(1+z_b\right)} & \bigg [   \left(\frac{z_a+\sigma _{aa}^{z_a} \sigma _{ab}}{1+z_a}\right){}^{-1+m_{aa}+m_{ba}} \left(\frac{z_b+\sigma _{ba} \sigma _{bb}^{z_b}}{1+z_b}\right){}^{-1+m_{ab}+m_{bb}} \bigg ] \cdot  \\ & \cdot \left(-z_b \sigma _{aa}^{z_a} \sigma _{ab} \left(-1+\sigma _{ba} \sigma _{bb}^{z_b}\right)+z_a \left(-1+\sigma _{aa}^{z_a} \sigma _{ab}\right) \left(-z_b+\left(-1+z_b\right) \sigma _{ba} \sigma _{bb}^{z_b}\right)\right).      \notag
\end{align*}

For arbitrary $z_a, z_b \in \mathbb{N}$, we obtain that the toppling branch generating functions are
\begin{align}
\mathcal{U}_a(\tau_a, \tau_b) &= \frac{(\tau_a + z_a)^{z_a} (\tau_b + z_b)}{(z_a+1)^{z_a} (z_b+1)},\label{UaRR11}\\
\mathcal{U}_b(\tau_a, \tau_b) &= \frac{(\tau_a + z_a) (\tau_b + z_b)^{z_b}}{(z_a+1) (z_b+1)^{z_b}}.\label{UbRR11}
\end{align}
Now the toppling size distribution can be read off the generating function obtained from Theorem \ref{theorem10}, which we compute symbolically in \emph{Mathematica}.

\subsection{Regular($z_a$)-Bernoulli($p$)-Regular($z_b$)}
One-to-one coupling is unsatisfactory for describing real world networks, in which nodes in one network may connect to arbitrarily many nodes in other networks (even none). A natural next choice for inter-network coupling is the Bernoulli distribution: each node has one neighbor in the other network (chosen uniformly at random) independently with probability $p$, no neighbor with probability $1-p$.  That is, the inter-degrees are Bernoulli distributed with parameter $p$. This allows variable coupling and no longer requires that the number of $a$-nodes equals the number of $b$-nodes. The distribution's finite support means that the generating functions have finitely many terms.

For simplicity, we let $p$ be the same for $a$ and $b$. Then, denoting the probability density function of the Bernoulli distribution by $B$, the degree distributions are
\begin{align*}
p_a(k_a, k_b) &= \delta_{z_a}(k_a) B(k_b; p),\\
p_b(k_a, k_b) &= B(k_a; p) \delta_{z_b}(k_b),
\end{align*}
which generate
\begin{align*}
G_a(\omega_a, \omega_b) &= (1-p) \omega_a^{z_a} + p \omega_a^{z_a} \omega_b,\\
G_b(\omega_a, \omega_b) &= (1-p) \omega_b^{z_b} + p \omega_a \omega_b^{z_b}.
\end{align*}
The branch distributions simplify to
\begin{align*}
q_{aa}(r_{aa}, r_{ab}) &= \delta_{z_a}(r_{aa}) \bigg [ \frac{p}{z_a+1} \delta_1(r_{ab}) + \frac{1-p}{z_a} \delta_0(r_{ab}) \bigg ],\\
q_{ab}(r_{ba}, r_{bb}) &= \frac{1}{z_b+1}  \delta_{z_b}(r_{bb}) \delta_1(r_{ba}),\\
q_{ba}(r_{aa}, r_{ab}) &= \frac {1} {z_a+1} \delta_{z_a}(r_{aa}) \delta_1(r_{ab}),\\
q_{bb}(r_{ba}, r_{bb}) &= \delta_{z_b}(r_{bb}) \bigg [ \frac{p}{z_b+1} \delta_1(r_{ba}) + \frac{1-p}{z_b} \delta_0(r_{ba}) \bigg ]
\end{align*}
away from the origin, and
\begin{align*}
q_{aa}(0,0) &= 1 - \frac{p}{z_a+1} - \frac{1-p}{z_a},\\
q_{ab}(0,0) &= \frac {z_b} {z_b + 1},\\
q_{ba}(0,0) &= \frac {z_a} {z_a + 1},\\
q_{bb}(0,0) &= 1 - \frac{p}{z_b+1} - \frac{1-p}{z_b}
\end{align*}
at the origin. Note that $q_{ab}, q_{ba}$ are identical to those for Regular-One-to-One-Regular, because a grain that is shed between the networks is equally likely to land on any of the nodes that have an inter-edge. The generating functions of the branch distributions are
\begin{align*}
\mathcal{Q}_{aa}(\sigma_{aa}, \sigma_{ab}) &= 1 - \frac{p}{z_a+1} - \frac{1-p}{z_a} + \frac {p}{z_a + 1} \sigma_{aa}^{z_a} \sigma_{ab}+ \frac {1-p}{z_a } \sigma_{aa}^{z_a},\\
\mathcal{Q}_{ab}(\sigma_{ba}, \sigma_{bb}) &= \frac {z_b}{z_b + 1} + \frac {1}{z_b + 1} \sigma_{ba} \sigma_{bb}^{z_b},\\
\mathcal{Q}_{ba}(\sigma_{aa}, \sigma_{ab}) &= \frac {z_a}{z_a + 1} + \frac {1}{z_a + 1} \sigma_{aa}^{z_a} \sigma_{ab},\\
\mathcal{Q}_{bb}(\sigma_{ba}, \sigma_{bb}) &= 1 - \frac{p}{z_b+1} - \frac{1-p}{z_b} + \frac {p}{z_b + 1} \sigma_{bb}^{z_b} \sigma_{ba}+ \frac {1-p}{z_b } \sigma_{bb}^{z_b}.
\end{align*}
The toppling branch generating functions are
\begin{align*}
\mathcal{U}_a(\tau_a, \tau_b) &= \frac{\big [ p(1 - \tau_a) + (z_a+1)(z_a + \tau_a - 1) \big ]^{z_a} (1 + z_b + p(\tau_b - 1))}{z_a^{z_a} (z_a+1)^{z_a} (1+z_b)},\\
\mathcal{U}_b(\tau_a, \tau_b) &= \frac{\big [ p(1 - \tau_b) + (z_b+1)(z_b + \tau_b - 1) \big ]^{z_b} (1 + z_a + p(\tau_a - 1))}{z_b^{z_b} (z_b+1)^{z_b} (1+z_a)}.
\end{align*}
When $p=1$, these reduce to the generating functions \eqref{UaRR11}, \eqref{UbRR11} for two regular graphs with one-to-one coupling.

\subsection{Current work on degree distributions with infinite support}
Currently we are working on other degree distributions that are not Kronecker delta functions nor Bernoulli probability density functions. The trouble with Poisson- and power-law-distributed degree distributions---and any others that have infinite support---is that the branch generating functions contain double-summations that are difficult (perhaps impossible) to simplify because of the coupling. For example, the shedding branch generating functions for two power-law networks with Poisson-distributed coupling are
\begin{align*}
\mathcal{Q}_{aa}(k_a, k_b) &= q_{aa}(0,0) + \sum_{k_a + k_b >0}  \frac{e^{-\lambda } \lambda ^{k_b} k_a^{1-\alpha }}{k_b! \left(k_a+k_b\right) \zeta (-1+\alpha )}   \omega_a^{k_a} \omega_b^{k_b},\\
\mathcal{Q}_{ab}(k_a, k_b) &= q_{ab}(0,0) + \sum_{k_a + k_b >0} \frac{e^{-\mu } \mu ^{-1+k_a} k_a k_b^{-\beta }}{k_a! \left(k_a+k_b\right) \zeta (\beta )}  \omega_a^{k_a} \omega_b^{k_b},\\
\mathcal{Q}_{ba}(k_a, k_b) &= q_{ba}(0,0) + \sum_{k_a + k_b >0} \frac{e^{-\lambda } \lambda ^{-1+k_b} k_a^{-\alpha } k_b}{k_b! \left(k_a+k_b\right) \zeta (\alpha )}   \omega_a^{k_a} \omega_b^{k_b},\\
\mathcal{Q}_{bb}(k_a, k_b) &= q_{bb}(0,0) + \sum_{k_a + k_b >0}  \frac{e^{-\mu } \mu ^{k_a} k_b^{1-\beta }}{k_a! \left(k_a+k_b\right) \zeta (-1+\beta )}   \omega_a^{k_a} \omega_b^{k_b},
\end{align*}
where $\zeta(\cdot)$ is the Riemann zeta function. Note that the factor $1 / (k_a + k_b)$ couples the double summation in a way that prevents us from separating the sums and analytically simplifying it. As a result, we cannot simplify the generating function to use a polylogarithm, as you can for an isolated power-law network \cite{goh_sandpile_powerlaw}. Nevertheless, the toppling branch generating functions may be analytically tractable, and of course one could always truncate sums before applying Theorem \ref{theorem10}.

\section{Results of generating function predictions and computer simulation}
First we compare the theoretical predictions of the generating function framework to computer simulations of the sandpiles. (Another way to match theory and experiment would be to simulate percolation with edge traversal probability $1/k$, where $k$ is the degree of the destination; this type of percolation may resemble processes other than cascading failures in infrastructure.) Next we show that the coupling of networks makes them less vulnerable to large avalanches, in contrast to the conclusion of the simpler model in \cite{NewmanHICSS04}. However, we also show that coupled networks suffer avalanches that are large in both networks more frequently than uncoupled networks. Finally we report on numerical simulations that are not yet covered by the mathematical framework, in which load is dropped on one network and we measure the size of avalanches in the other network, which further illustrates the \emph{destabilizing} effect of coupling between networks.

\subsection{Regular(3)-One-to-One-Regular(10): matching theory and experiment}
Here we compare the theoretical predictions of avalanche size to a simulation with two regular graphs $a, b$ with uniform internal degrees $z_a = 3, z_b = 10$ and one-to-one coupling. The parameters are $N_a = N_b = 10^4$ nodes per network, where each node is initialized with a number of grains chosen uniformly from 0 to a node's degree minus one (to expedite the simulation) and we initially run $10^4$ ``transient'' events (grains for which we do not record statistics).  We then record statistics for an addition $10^5$ grains of sand dropped, using dissipation rate $f = 0.001$.

\begin{figure}[hbt]
\begin{center}
\includegraphics[width=16cm]{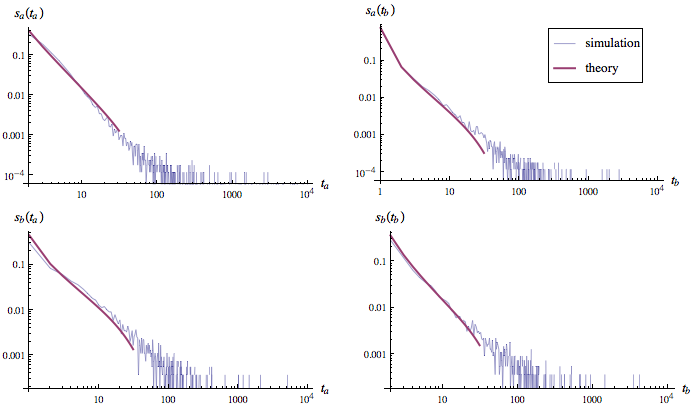}
\caption{Comparison of theoretical predictions (red line) with simulation (blue line) of avalanche size distributions for Regular(3)-One-to-One-Regular(10). We denote $s_a(t_a) := \sum_{t_b=0}^\infty s_a(t_a, t_b)$ 
(and similarly for $s_a(t_b), s_b(t_a), s_b(t_b)$). For the theoretical predictions, we computed 40 terms in Eq. \eqref{eq:theorem10}, which generate 8000 terms for each generating function.}
\label{simulation_theory_RR11_310}
\end{center}
\end{figure}

The match between simulation and theory is good. As predicted by percolation theory (e.g., \cite{durrett_randomgraphdynamics}),  the branching process approximation of a random graph of $n$ nodes holds well for $O(\sqrt{n})$ many nodes. That is, after an avalanche affects approximately $O(\sqrt{n})$ many nodes, the ``branches'' of the ``avalanche tree'' begin to collide due to the presence of large loops in the graph. We see that effect in our simulations: for $10^4$ nodes, the avalanche is tree-like up to about $10^2$ nodes.

\subsection{Coupling between networks can stabilize them individually}\label{Section:stabilizing}
\begin{figure}[hbt]
\begin{center}
\includegraphics[width=6.38in]{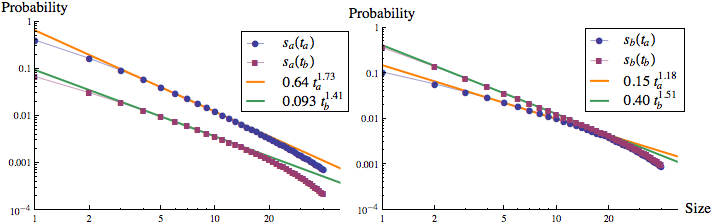}
\caption{Avalanche size distributions within the networks are not power laws. Plotted in log-log scales in blue and maroon are the generating function predictions of avalanche size distributions $s_a(t_a, t_b)$ (left) and $s_b(t_a, t_b)$ (right) for Regular(3)-One-to-One-Regular(10). Plotted in orange and green are best fit lines to data points 3 through 15, which show that the avalanche distributions are sublinear for large avalanches due to sand leaking across to the other network (not due to finite size effects). We denote $s_a(t_a) := \sum_{t_b=0}^\infty s_a(t_a, t_b)$ (and similarly for $s_a(t_b), s_b(t_a), s_b(t_b)$). Here we computed 40 summands in Eq. \eqref{eq:theorem10}.}
\label{s_a_s_b}
\end{center}
\end{figure}

What is interesting about these cascades in interacting networks in Fig. \ref{simulation_theory_RR11_310} is that, unlike most variants of the sandpile model, the avalanche size distributions are not quite power laws. In Fig. \ref{s_a_s_b} we plot the generating function predictions with the lines showing the best fits when data points 3 through 15 are considered, illustrating that the avalanche size distributions fall short of being power laws at large avalanche size. This is not due to finite size effects, because the generating function predictions are independent of the number of nodes. Instead the avalanches fall short of being a power law for large avalanche size because sand leaks to the other network, an effect which is heightened for large avalanches. 

\begin{figure}[hbt]
\begin{center}
\includegraphics[width=12cm]{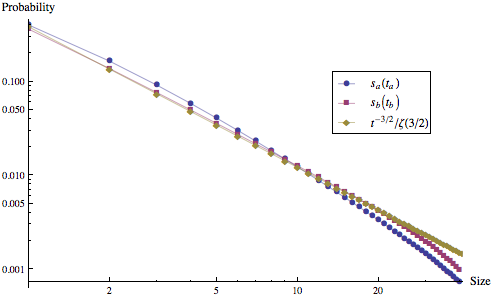}
\caption{In networks sparsely connected to others, large avalanches are mitigated and small avalanches are amplified compared to the mean-field behavior that the network would exhibit if it were isolated (power-law with exponent 3/2), due to the stabilizing effect of connecting networks. Here we plot in blue circles and maroon squares the generating function predictions for the marginalized avalanche distributions $s_a(t_a), s_b(t_b)$ (defined in Fig. \ref{s_a_s_b}'s caption) for Regular(3)-One-to-One-Regular(10) coupled networks, together with the normalized power-law with exponent 3/2 (gold diamonds).} 
\label{sub_mean-field}
\end{center}
\end{figure}

As illustrated in Fig. \ref{sub_mean-field}, when compared to mean-field behavior (avalanche size distribution a power law with exponent $3/2$), in coupled networks the avalanche size distributions within each network have more frequent small avalanches and less frequent large avalanches. Yet, the avalanches in the flavorless view (Null Hypothesis \ref{flavorlessview})
---i.e., the two interacting networks viewed as one network without flavors---do show the mean-field behavior, consistent with the robustness of the mean-field behavior over different network structures \cite{goh_sandpile_powerlaw}. Nonetheless we show that avalanches within a network that is coupled to another are less likely to be large than one would predict if the network were isolated.

\begin{figure}[hbt]
\begin{center}
\includegraphics[width=12cm]{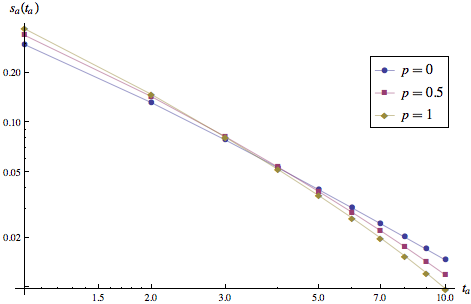}
\caption{Strengthening the coupling between two networks further mitigates large avalanches. Here we show the generating function predictions of the avalanche size distribution $s_a(t_a) \equiv \sum_{t_b = 0}^\infty s_a(t_a, t_b)$ of avalanches begun in network $a$ for two random regular graphs, each with uniform intra-degree 3 and connected via Bernoulli-distributed coupling with parameter $p$ (shown here for $p = 0$, $p=0.5$ and $p=1$). As $p$ increases, large avalanches become less likely and small avalanches become more likely, due to the stabilizing effect of the coupling to the other network.} 
\label{Bernoulli_varyp}
\end{center}
\end{figure}

This stabilizing effect strengthens with increased coupling between networks. We explore this using the generating function predictions for two random regular graphs with Bernoulli-distributed coupling (i.e., each node has a neighbor in the other network with probability $p$). As illustrated in Fig. \ref{Bernoulli_varyp}, as we strengthen the coupling between the networks (i.e., the parameter $p$ of the Bernoulli distribution), large avalanches become progressively less likely while small avalanches become more likely. In Fig. \ref{Bernoulli_varyp} networks $a$ and $b$ are random 3-regular graphs (internally); we choose the same intra-degree for both networks in order to isolate the effect of the coupling, though the same conclusion---that increased coupling between the networks mitigates large avalanches and amplifies small ones---holds for random regular graphs with different intra-degrees, as well.

These two effects---that large avalanches are less likely 
and are further suppressed with increasing coupling between networks---suggests that coupling a network to other networks stabilizes it, because the other networks act as reservoirs for extra load. This can be understood intuitively as follows. Suppose an isolated network is poised to be overwhelmed by an avalanche, with many nodes dangerously near their capacities. Coupling the network to another network may mitigate young avalanches since it sheds some of its load during its crucial, early formation, before it grows sufficient size and momentum to overwhelm the whole network. Moreover, avalanches that leak across to another network are likely to be small there, and only rarely do they grow large and hence become likely to amplify the cascade in the original network. This stabilizing effect of coupling networks contrasts to the result for a simpler version of this model in \cite{NewmanHICSS04}, where they concluded that coupling networks always destabilizes. The assumption in \cite{NewmanHICSS04} that nodes shed sand to \emph{every} node means that load shed to the other network can only exacerbate the cascade; 
here, load shed to the other network frequently mitigates the cascade, and only rarely exacerbates it.

This result also ameliorates the warnings in \cite{buldyrev} about the catastrophic cascades of disrupted connectivity in coupled networks. Buldyrev et al.\ found that coupling two networks exacerbates cascades of failing connectivity because coupling the networks can only provide new ways to fail and never ways to mitigate them, whereas in this model the coupling can both stabilize and destabilize a network. Yet there are regimes where introducing connections between networks can destabilize them as shown next. 

\subsection{Coupling networks can destabilize them jointly}
Although large avalanches are mitigated in networks that are connected to other networks (compared to isolated networks), when large avalanches do occur in a network they more frequently accompany large avalanches in the other network. The intuition is clear: a large cascade in one network likely leaks across the weak coupling to the other network, and the cascades in the two networks amplify one another. Said differently, large avalanches in the \emph{marginalized} avalanche size distributions are reduced, but in the \emph{joint} avalanche size distribution for the two networks, avalanches large in both networks become more likely than for isolated networks.

As in Section \ref{Section:stabilizing}, we compare the joint avalanche size distribution for two coupled networks to the null hypothesis of two isolated networks. For two isolated, random regular graphs, the avalanche size distributions are mean-field, so their joint avalanche size distribution is the product measure of two power-laws with exponent $3/2$,
\begin{align}\label{joint_uncoupled}
s^\text{uncoupled}(t_a, t_b) = \frac {(t_a t_b)^{-3/2}} {\zeta(3/2)^2}
\end{align}
where $\zeta$ is the Riemann zeta function. For two \emph{coupled} networks, the appropriate joint distribution to compare to Eq. \eqref{joint_uncoupled} is
\begin{align}\label{joint_coupled}
s(t_a, t_b) = \frac{1}{2} (s_a(t_a, t_b) + s_b(t_a, t_b)).
\end{align}
The justification of Eq. \eqref{joint_coupled} is as follows. We drop grains of sand uniformly at random on the nodes, and there are as many $a$-nodes as $b$-nodes, so the chance that the sand lands on an $a$-node or $b$-node is a fair coin toss ($1/2, 1/2$). Conditioned on where it lands, the chance that the first node topples is roughly the same for both networks, because we find numerically that grains are approximately uniformly distributed from 0 to $k-1$, where $k$ is the degree of the node. Conditioned on these two events, the chance that the ensuing avalanche topples $t_a$ many $a$-nodes and $t_b$ many $b$-nodes is $s_a(t_a, t_b)$ or $s_b(t_a, t_b)$, respectively.

\begin{figure}[hbt]
\begin{center}
\includegraphics[width=9cm]{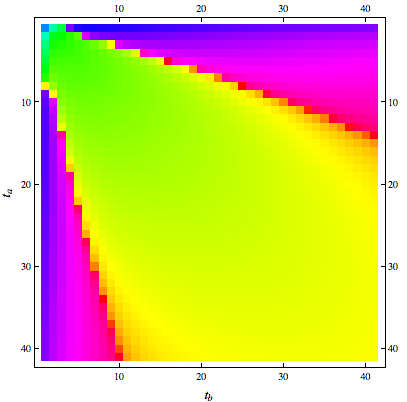} \hspace{1mm} \includegraphics[width=1.7cm]{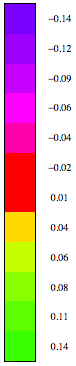}
\caption{Avalanches that are large in both networks (yellow area, bottom-right) are more likely in coupled networks than in uncoupled networks. Plotted in color is $\mathcal{L} [s(t_a, t_b) - s^\text{uncoupled}(t_a, t_b)]$, where $\mathcal{L}(x) := \text{sgn}(x) \log(|x|)$, and where $s(t_a, t_b)$ and $s^\text{uncoupled}(t_a, t_b)$ are the joint avalanche size distributions for two random 3-regular and 10-regular graphs 
 with one-to-one coupling and with no coupling, respectively, defined in Eqs. \eqref{joint_coupled} and \eqref{joint_uncoupled}. Yellow and green correspond to $s > s^\text{uncoupled}$; red indicates that avalanches of those sizes are equally likely in coupled and uncoupled networks; and purple and yellow indicate that  $s < s^\text{uncoupled}$. Values in the color legend are $s(t_a, t_b) - s^\text{uncoupled}(t_a, t_b)$.}
\label{compare_joints}
\end{center}
\end{figure}

Fig. \ref{compare_joints} compares the joint distribution \eqref{joint_coupled} of two coupled networks with the joint distribution \eqref{joint_uncoupled} of two uncoupled networks. The large sea of yellow in the bottom-right corner indicates that $s(t_a, t_b) > s^\text{uncoupled}(t_a, t_b)$ for $t_a, t_b$ both large---i.e., that avalanches large in one network more frequently accompany avalanches large in the other network in coupled networks compared to uncoupled networks.

\subsection{Numerical experiment: dropping grains only in network $a$}
The multi-type branching approximation above shows that sandpile models on two weakly coupled networks rather than on an isolated network reduces the frequency of large avalanches. However, dropping sands in only one of the two networks (rather than in both) magnifies the distinction between coupled and isolated networks. Here we consider dropping sand only in network $a$, though one could use other rules such as dropping on average two grains in $a$ for every one grain dropped in $b$.

Dropping sand only in network $a$ pushes $a$-nodes to their capacities, thereby causing many avalanches in $a$ that occasionally leak to $b$ via the sparse coupling. On one hand, network $b$ can stabilize network $a$ by serving as a reservoir for dumping extra load. On the other hand, avalanches in $b$ can explode in size and thus amplify avalanches in $a$. Which effect is more pronounced asymptotically?

\begin{figure}[htb]
\begin{center}
\includegraphics[width=16cm]{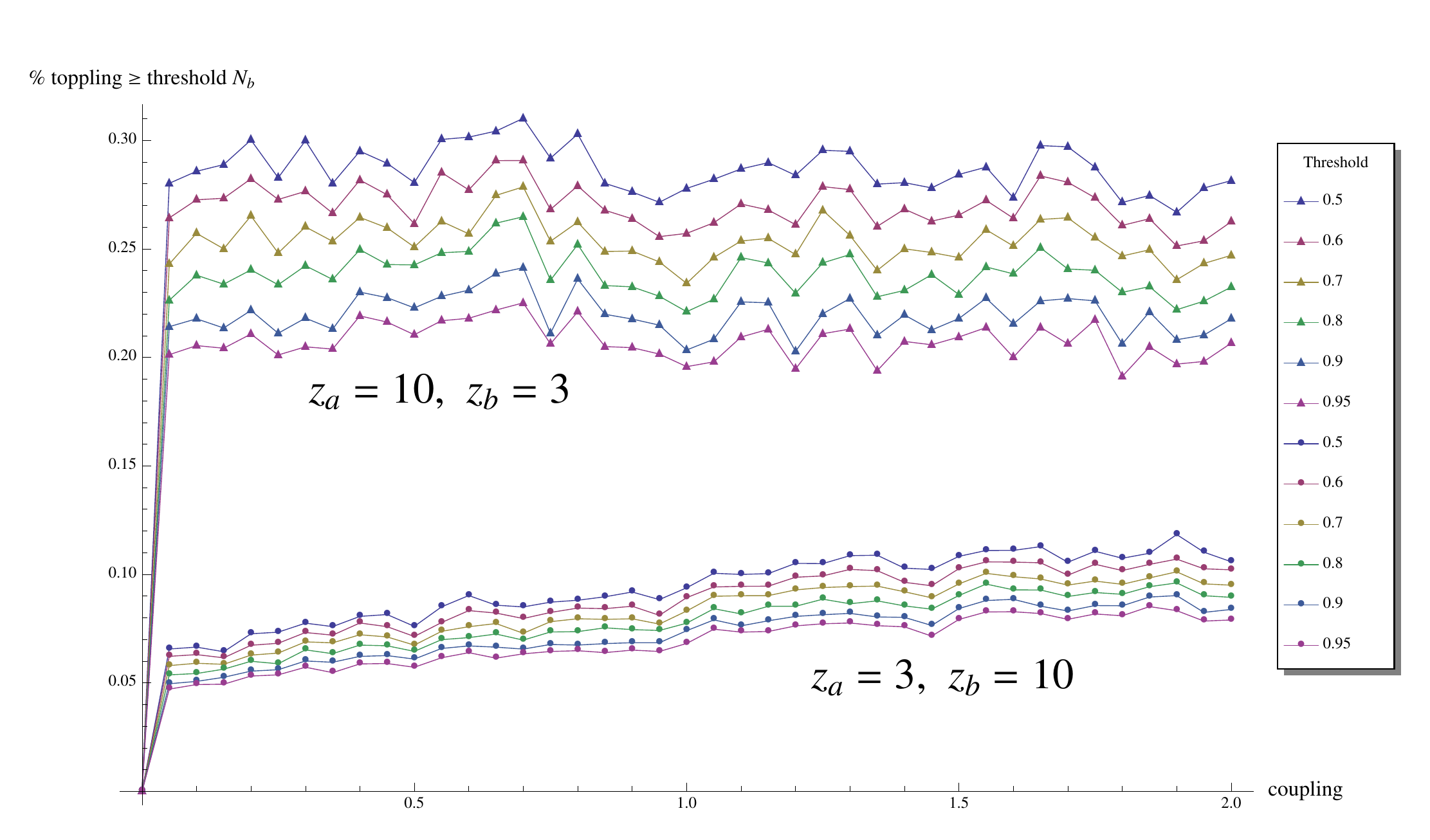}
\caption{With sand dropped in network $a$ uniformly at random, the avalanches in network $b$ are larger when it is less dense than $a$ (triangles) compared to when it is more dense than $a$ (circles). We vary the coupling between the networks by tuning the means of the Poisson inter-degree distributions from 0.01 to 2.0 (horizontal axis). Plotted vertically is the fraction of avalanches that topple $\geq t N_b$, where the threshold $t$ is varied from 0.5 (blue) to 0.95 (gold), and $N_b$ is the number of $b$-nodes. The networks are regular graphs with $N_a = N_b = 1000$ nodes in each network and uniform intra-degree $z_a = 10, z_b = 3$ (triangles, above), $z_a = 3, z_b = 10$ (circles, below).}
\label{varycoupling}
\end{center}
\end{figure}

First we explore how the size and frequency of large avalanches in network $b$ depend on the strength of the coupling between the networks. In Fig. \ref{varycoupling} we plot the fraction of avalanches that topple a certain large fraction of $b$-nodes as a function of the coupling strength, which here is the mean of the two Poisson inter-degree distributions. There is no phase transition at some positive critical coupling strength: once the networks are even slightly coupled (Poisson mean 0.01), avalanches in $a$ can overwhelm $b$. That is, a tiny capacity to leak avalanches across the networks suffices to topple nearly all of network $b$ roughly every twentieth avalanche (when $z_a=3, z_b=10$).

Second, we explore how the severity of avalanches in $b$ depend on the relative density of intra-edges in $a$ and $b$. Although the fraction of large avalanches in $b$ saturates once the networks are slightly coupled, the size and frequency of large avalanches in $b$ depends on which network is more dense (i.e., has more intra-edges). As illustrated in Fig. \ref{varycoupling}, when $a$ and $b$ are random regular graphs with weak Poisson coupling, the large avalanches in $b$ (the network not receiving external load) are \emph{more frequent and larger} when $b$ is less dense (uniform degree $z_b = 3 < z_a=10$, triangles in Fig. \ref{varycoupling}) compared to when $b$ is more dense (uniform degree $z_b = 10 > z_a=3$, circles in Fig. \ref{varycoupling}).

Intuitively, when $a$ is more dense than $b$, $a$ has more total capacity---since here capacities are degrees---so the large cascades in $a$ consist of so much sand, relative to the total capacity of $b$, that it can easily overwhelm $b$ by jumping across the weak coupling. As a result, to prevent a network from large cascades caused by a neighboring network, one should increase the network's capacity to be comparable to or larger than the neighboring network's capacity, which in this model means add more edges since capacities are the nodes' degrees.


\section{Discussion and Conclusion}
We develop a mathematical framework for interdependent networks and for approximating cascades on them using multi-type branching processes. Along the way we elucidate the bias induced by requiring matching undirected edge stubs in bipartite and interacting networks. Using a generalization of Lagrange's expansion to several variables \cite{good_lagrange}, we solve the branching process equations for the avalanche size distributions for random regular graphs with one-to-one or with Bernoulli-distributed coupling, and show the theoretical predictions match well with results from simulations. For sandpile models, we find that coupled networks are more stable than isolated ones, in that large avalanches occur less frequently due to sand leaking across to the other network, in contrast to the conclusion in \cite{NewmanHICSS04} using a simpler sandpile model and to the conclusion in \cite{buldyrev} for a model of cascading failures of connectivity in coupled networks. However, we also show that coupling networks can destabilize them, in that large avalanches, though more rare, more frequently accompany large avalanches in the neighboring network. Furthermore, disparity in capacity and in applied load can enhance large avalanches: when load is applied to one network, large avalanches in the second network increase in severity and in frequency, an effect that is amplified with increased coupling between the networks and with increased disparity in relative capacity.

These findings suggest economic and game-theoretic implications for infrastructure. On one hand, a greedy owner of an electrical grid (say) prefers to add connections to other networks, since this mitigates large cascades in her own network. By contrast, to mitigate avalanches that are simultaneously large in the entire collection of connected electrical grids, the grids should be \emph{less} coupled to one another. Thus, like in the game Prisoner's dilemma \cite{PrisonersDilemma}, the action optimal for societal welfare---\emph{decrease coupling between networks}---conflicts with the action optimal for individual networks---\emph{increase coupling between networks}. More detailed economic models that combine results like those here with economic and physical considerations of electrical grids may elucidate what networks optimally balance the stability to large avalanches with the cost of adding connectivity.

Here we have focused on mitigating large avalanches, with the example of cascading failures in infrastructure in mind. However, cascades in coupled networks apply equally well to situations in which we wish to \emph{enhance} large avalanches. For example, advertisers who design word-of-mouth campaigns to spread adoption of products in social networks \cite{BanerjeeFudenbergWordofMouth} wish it to spread from one sub-population to another across sparse connections. Meanwhile, sociologists who use response-driven surveys (RDS) to collect data---in which they pay subjects to recruit friends to participate in the survey---wish that their surveys penetrate bottlenecks to spread to different populations (say, from drug users to homosexuals) \cite{SalganikHeckathornRDS}. To understand cascades in interacting networks in social contexts such as these would require different branch distributions for the branching process (rather than ones inversely proportional to degree, as for the sandpile model here) and networks with clustering or with arbitrary subgraphs \cite{newman_subgraphs, miller_percolation, newman_clustering}. We expect that the tools for solving multiplicative branching processes developed by I. J. Good \cite{good_lagrange} will find other uses for locally tree-like networks, as it is straightforward to implement using computer algebra systems; to study cascades on interacting networks that are not tree-like would require modification (along the lines of \cite{newman_subgraphs, miller_percolation, newman_clustering}).

Yet there is more work do for the cascades on tree-like graphs considered here, in particular to more faithfully capture infrastructure. For instance, different rules for shedding load other than ``shed one grain of sand to each neighbor'' deserve attention (see, e.g., \cite{goh_sandpile_powerlaw_supplement}). An example shedding rule for interacting networks is that nodes preferentially shed to their network or to other networks. Another open problem is to combine models of bearing and shedding load, such as sandpiles, with models of cascading failures of connectivity, such as the model in \cite{buldyrev}; as we have shown, these two models yield contrasting conclusions regarding the effect that coupling between networks has on their stability, so it would be interesting to study which effect dominates.

More broadly, this work suggests that using knowledge about the connection structure within and between modules in networks---or alternatively about the structure within and between disparate networks---can help to predict processes on them\footnote{Perhaps a new criterion for choosing the best partitions of a network into modules should be, ``How well does it let me predict dynamical processes on it?''}, be it cascades, contact processes, synchronization, or other dynamics.

\begin{acknowledgements}
CB was supported by NSF VIGRE DMS0636297. We also gratefully acknowledge support from the National Academies Keck Futures Initiative under Grant No. CS05 and support from the Defense Threat Reduction Agency, Basic Research Award \#DTRA1-10-1-0088.
\end{acknowledgements}

\bibliographystyle{siam}
\bibliography{networksbib}

\end{document}